\documentclass{article}
\usepackage{graphicx}
\usepackage{url}
\usepackage{color}
\usepackage{graphicx}
\usepackage[font=small]{caption}
\usepackage{natbib}
\usepackage{amsmath}
\usepackage{mathtools}
\usepackage{multirow}
\usepackage{algpseudocode,algorithm,algorithmicx}
\usepackage{hyperref,subfigure,pgf,fullpage,url,tikz,afterpage,placeins,amsmath,bm,bbm,longtable,amssymb,listings}
\usepackage{multirow}
\usepackage{booktabs}
\usepackage{authblk}
\usepackage{pdfpages}

\usepackage{titlesec}

\newcommand*\Let[2]{\State #1 $\gets$ #2}
\algrenewcommand\algorithmicrequire{\textbf{Input:}}
\algrenewcommand\algorithmicensure{\textbf{Output:}}

\begin{document}

\title{Modeling outcomes of soccer matches}


\author[1]{Alkeos Tsokos}
\author[2]{Santhosh Narayanan}
\author[2,3]{Ioannis Kosmidis}
\author[1]{Gianluca Baio}
\author[3,4]{Mihai Cucuringu}
\author[1]{Gavin Whitaker}
\author[1,3]{Franz Kir\'aly}
\affil[1]{University College London}
\affil[2]{University of Warwick}
\affil[3]{The Alan Turing Institute}
\affil[4]{University of Oxford}


\date{\today}

\maketitle

\begin{abstract}
  \noindent We compare various extensions of the Bradley-Terry model and a
  hierarchical Poisson log-linear model in terms of their performance
  in predicting the outcome of soccer matches (win, draw, or
  loss). The parameters of the Bradley-Terry extensions are estimated
  by maximizing the log-likelihood, or an appropriately penalized
  version of it, while the posterior densities of the parameters of
  the hierarchical Poisson log-linear model are approximated using
  integrated nested Laplace approximations. The prediction performance
  of the various modeling approaches is assessed using a novel,
  context-specific framework for temporal validation that is found to
  deliver accurate estimates of the test error. The direct modeling
  of outcomes via the various Bradley-Terry extensions and the
  modeling of match scores using the hierarchical Poisson log-linear
  model demonstrate similar behavior in terms of predictive
  performance.   \vspace{0.3cm} 
\end{abstract}

\section{Introduction}
The current paper stems from our participation in the 2017 Machine
Learning Journal (Springer) challenge on predicting outcomes of soccer
matches from a range of leagues around the world (MLS challenge, in
short). Details of the challenge and the data can be found in
\citet{Berrar2017}.

We consider two distinct modeling approaches for the task. The first
approach focuses on modeling the probabilities of win, draw, or loss,
using various extensions of Bradley-Terry models
\citep{Bradley1952}. The second approach focuses on directly modeling
the number of goals scored by each team in each match using a
hierarchical Poisson log-linear model, building on the modeling
frameworks in \citet{Maher1982}, \citet{Dixon1997}, \citet{Karlis2003}
and \citet{Baio2010}.

The performance of the various modeling approaches in predicting the
outcomes of matches is assessed using a novel, context-specific framework for
temporal validation that is found to deliver accurate estimates of the
prediction error. The direct modeling of the outcomes using the
various Bradley-Terry extensions and the modeling of match scores
using the hierarchical Poisson log-linear model deliver similar
performance in terms of predicting the outcome.

The paper is structured as follows: Section \ref{data} briefly
introduces the data, presents the necessary data-cleaning operations
undertaken, and describes the various features that were
extracted. Section \ref{bt} presents the various Bradley-Terry models
and extensions we consider for the challenge and describes the
associated estimation procedures. Section \ref{poisson} focuses on the
hierarchical Poisson log-linear model and the Integrated Nested
Laplace Approximations (INLA; \citealt{Rue2009}) of the posterior
densities for the model parameters. Section \ref{validationsec}
introduces the validation framework and the models are compared in
terms of their predictive performance in Section
\ref{results}. Section~\ref{conclusion} concludes with discussion and
future directions.

\section{Pre-processing and feature extraction} \label{data}

\subsection{Data exploration}
\label{features}
The data contain matches from $52$ leagues, covering $35$ countries,
for a varying number of seasons for each league. Nearly all leagues
have data since 2008, with a few having data extending as far back as
2000. There are no cross-country leagues (e.g.~UEFA Champions League)
or teams associated with different countries. The only way that teams move between leagues is within each country by either promotion or relegation.
\begin{figure}[t]
\centering
\includegraphics[scale=0.5]{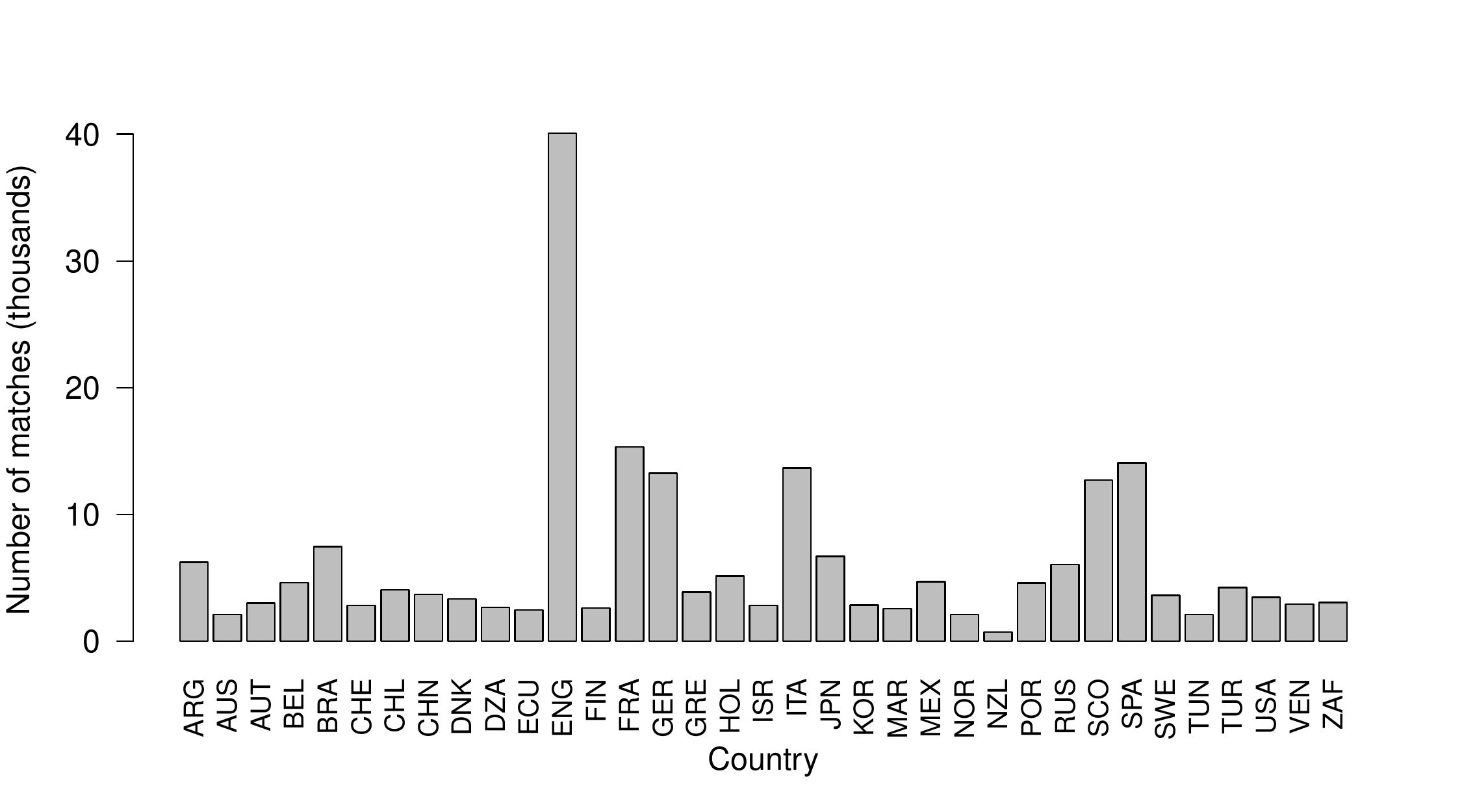}
\caption{Number of available matches per country in the
  data. 
}
\label{fig-country}
\end{figure}

Figure~\ref{fig-country} shows the number of available matches for
each country in the data set. England dominates the data in terms of
matches recorded, with the available matches coming from 5 distinct
leagues. The other highly-represented countries are Scotland with data
from 4 distinct leagues, and European countries, such as Spain,
Germany, Italy and France, most probably because they also have a high
UEFA coefficient \citep{wiki:UEFA_coefficient}.

\begin{figure}[t]
\centering
\includegraphics[scale=0.5]{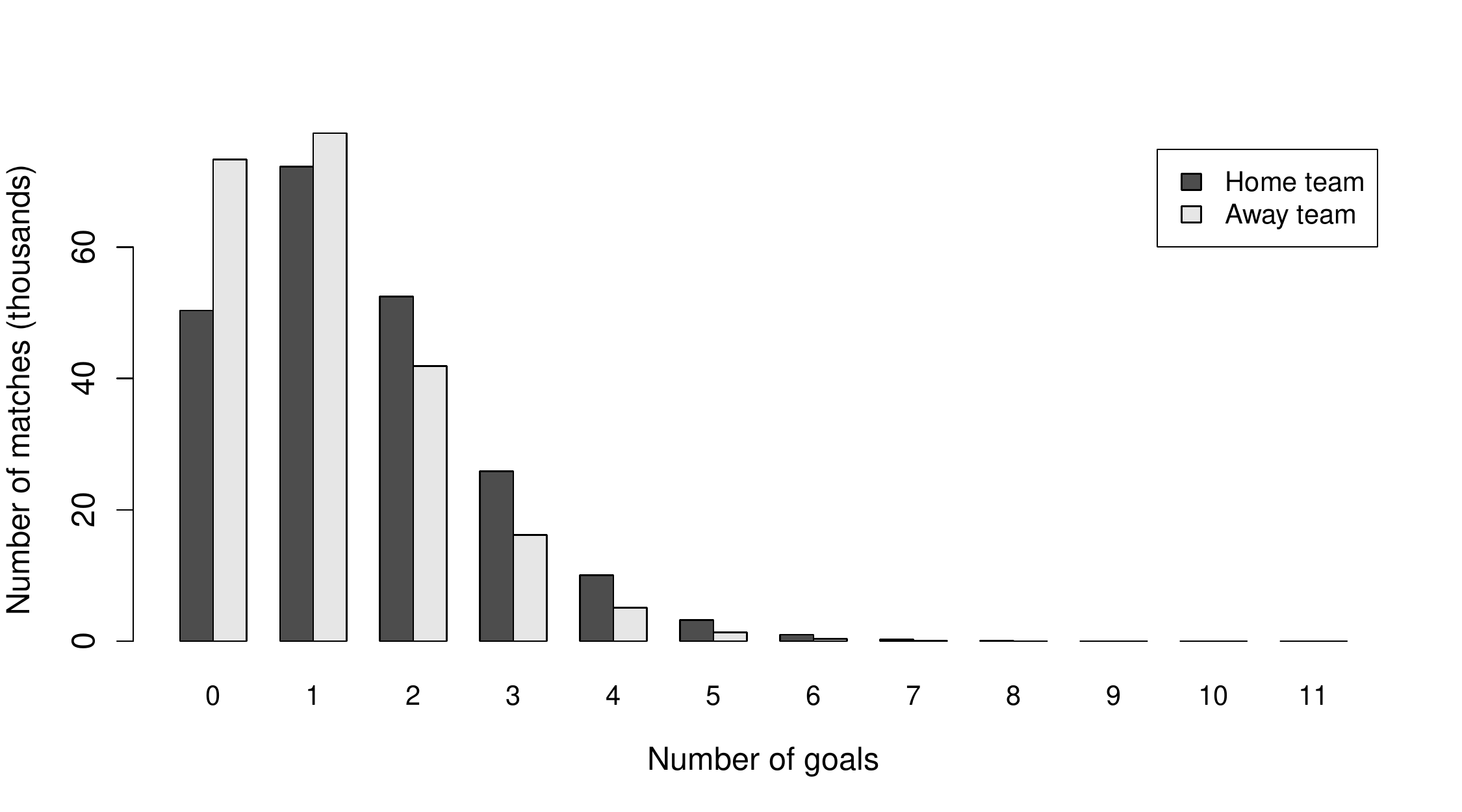}
\caption{The number of matches per number of goals scored by the home
  (dark grey) and away team (light grey). 
}
\label{fig-goals}
\end{figure}

Figure~\ref{fig-goals} shows the number of matches per number of goals
scored by the home (dark grey) and away (light grey) teams. Home teams
appear to score more goals than away teams, with home teams having
consistently higher frequencies for two or more goals and away teams
having higher frequencies for no goal and one goal. Overall, home
teams scored $304,918$ goals over the whole data set, whereas away
teams scored $228,293$ goals. In Section 1 of the Supplementary
Material, the trend shown in Figure~\ref{fig-goals} is also found to
be present within each country, pointing towards the existence of a
home advantage.

\subsection{Data cleaning}
Upon closer inspection of the original sequence of matches for the MLS
challenge, we found and corrected the following three anomalies in the
data. The complete set of matches from the 2015-2016 season of the
Venezuelan league was duplicated in the data. We kept only one
instance of these matches. Furthermore, $26$ matches from the
2013-2014 season of the Norwegian league were assigned the year 2014
in the date field instead of 2013. The dates for these matches were
modified accordingly. Finally, one match in the 2013-2014 season of
the Moroccan league (Raja Casablanca vs Maghrib de Fes) was assigned
the month February in the date field instead of August. The date for
this match was corrected, accordingly.

\subsection{Feature extraction}
The features that were extracted can be categorized into team-specific,
match-specific and/or season-specific. Match-specific features were
derived from the information available on each match. Season-specific
features have the same value for all matches and teams in a season of
a particular league, and differ only across seasons for the same
league and across leagues.

Table~\ref{tab-features} gives short names, descriptions, and ranges for
the features that were extracted. Table~\ref{tab-artificial_features}
gives an example of what values the features take for an artificial
data set with observations on the first 3 matches of a season for team A
playing all matches at home. The team-specific features are listed
only for Team A to allow for easy tracking of their evolution.

The features we extracted are proxies for a range of aspects of the
game, and their choice was based on common sense and our
understanding of what is important in soccer, and previous
literature. Home (feature 1 in Table~\ref{tab-artificial_features})
can be used for including a home advantage in the models; newly
promoted (feature 2 in Table~\ref{tab-artificial_features}) is used to
account for the fact that a newly promoted team is typically weaker
than the competition; days since previous match (feature 3 in
Table~\ref{tab-artificial_features}) carries information regarding
fatigue of the players and the team, overall; form (feature 4 in
Table~\ref{tab-artificial_features}) is a proxy for whether a team is
doing better or worse during a particular period in time compared to its
general strength; matches played (feature 5 in
Table~\ref{tab-artificial_features}) determines how far into the
season a game occurs; points tally, goal difference, and points per
match (features 6, 7 and 10 in Table~\ref{tab-artificial_features})
are measures of how well a team is doing so far in the season; goals
scored per match and goals conceded per match (features 8 and 9 in
Table~\ref{tab-artificial_features}) are measures of a team's
attacking and defensive ability, respectively; previous season points
tally and previous season goal difference (features 11 and 12 in
Table~\ref{tab-artificial_features}) are measures of how well a team
performed in the previous season, which can be a useful indicator of
how well a team will perform in the early stages of a season when
other features such as points tally do not carry much information;
finally, team rankings (feature 13 in
Table~\ref{tab-artificial_features}) refers to a variety of measures
that rank teams based on their performance in previous matches, as
detailed in Section 2 of the Supplementary Material.

In order to avoid missing data in the features we extracted, we made
the following conventions. The value of form for the first match of
the season for each team was drawn from a Uniform distribution in
$(0,1)$. The form for the second and third match were a third of the
points in the first match, and a sixth of the total points in the
first two matches, respectively. Days since previous match was left
unspecified for the very first match of the team in the data. If the
team was playing its first season then we treated it as being newly
promoted. The previous season points tally was set to $15$ for newly
promoted teams and to $65$ for newly relegated teams, and the previous
season goal difference was set to $-35$ for newly promoted teams and
$35$ for newly relegated teams. These values were set in an ad-hoc
manner prior to estimation and validation, based on our sense and
experience of what is a small or large value for the corresponding
features. In principle, the choice of these values could be made more
formally by minimizing a criterion of predictive quality, but we did
not pursue this as it would complicate the estimation-prediction
workflow described later in the paper and increase computational
effort significantly without any guarantee of improving the predictive
quality of the models.

\def\arraystretch{1.5}
\begin{table*}[t]
  \caption{Short names, descriptions, and ranges for the features that
    were extracted.}
\scriptsize
  \begin{center}
    \begin{tabular}{llll}
      \toprule
      Number & Short name & Description & Range \\ \midrule
      \multicolumn{4}{l}{Team-specific features} \\ \midrule
      1 & Home &     \noindent\parbox[t]{7cm}{$1$ if the team is playing at home, and $0$ otherwise} & $\{0, 1\}$ \\
      2 & Newly promoted & \noindent\parbox[t]{7cm}{$1$ if the team is newly
                           promoted to the league for the current
                           season, and $0$ otherwise} & $\{0, 1\}$ \\
      3 & Days since previous match & \noindent\parbox[t]{7cm}{number of days elapsed since the previous
                                  match of the team} & $\{1, 2, \ldots\}$
      \\
      4 & Form & \noindent\parbox[t]{7cm}{a ninth of the total points gained in the last three
                 matches in the current season} & $(0, 1)$  \\
      5 & Matches played & \noindent\parbox[t]{7cm}{number of matches played in the current season and
                          before the current match} & $\{1, 2, \ldots\}$
      \\
      6 & Points tally & \noindent\parbox[t]{7cm}{the points accumulated during the current
                         season and before the current match} & $\{0, 1, \ldots\}$
      \\
      7 & Goal difference & \noindent\parbox[t]{7cm}{the goals that a team has scored minus the goals that it has conceded
                            over the current season and before the current match} & $\{\ldots,
                                                                                   -1, 0, 1,
                                                                                   \ldots\}$
      \\
      8 & Goals scored per match & \noindent\parbox[t]{7cm}{total goals scored per match played over
                                   the current season and before the current match} & $\Re^{+}$ \\
      9 & Goals conceded per match & \noindent\parbox[t]{7cm}{total goals conceded per match over the
                                     current season and before the current match} & $\Re^{+}$ \\
      10 & Points per match & \noindent\parbox[t]{7cm}{total points gained per match played over the
                              current season and before the current match} & $[0, 3]$ \\
      11 & Previous season points tally & \noindent\parbox[t]{7cm}{total points accumulated by the
                                      team in the previous season of
                                      the same league} & $\{0, 1, \ldots\}$ \\
      12 & Previous season goal difference & \noindent\parbox[t]{7cm}{total goals scored minus total goals conceded for each team in the previous season of the same league} & $\{\ldots,
                                                                                   -1, 0, 1,
                                                                                   \ldots\}$ \\
      13 & Team rankings & \noindent\parbox[t]{7cm}{a variety of team rankings, based on
                           historical observations; See Section~2 of the Supplementary Material}
                                        & $\Re$ \\ \midrule
      \multicolumn{4}{l}{Season-specific features} \\ \midrule
      14 & Season & \noindent\parbox[t]{7cm}{the league season in which each match is played} &
                                                                                                labels \\
      15 & Season window & \noindent\parbox[t]{7cm}{time period in calendar months of the
                           league season} & labels

      \\ \midrule
      \multicolumn{4}{l}{Match-specific features} \\ \midrule
      16 & Quarter & \noindent\parbox[t]{7cm}{quarter of the calendar year based on the match date} &
                                                                       labels \\
                                                                       \bottomrule
    \end{tabular}
  \end{center}
  \label{tab-features}
\end{table*}

\def\arraystretch{1}
\begin{table*}[t]
  \caption{Feature values for artificial data showing the first 3
    matches of a season with team A playing all matches at home.}
\scriptsize 
  \begin{center}
\begin{tabular}[t]{llll}
\toprule
  & Match 1     & Match 2     & Match 3
  \\  \cmidrule{2-4}
\multicolumn{4}{l}{Match attributes and outcomes} \\
\midrule
League                         & Country1       & Country1       &
                                                                   Country1 \\
Date                            & 2033-08-18 & 2033-08-21 & 2033-08-26 \\
Home team                   & team A     & team A     & team A     \\
Away team                   & team B     & team C     & team D     \\
Home score                 & 2          & 2          & 0          \\
Away score                 & 0          & 1          & 0          \\ \midrule
\multicolumn{4}{l}{Team-specific features (Team A)} \\ \midrule
Newly promoted                & 0          & 0          & 0          \\
Days since previous match         & 91         & 3          & 5          \\
Form                          & 0.5233     & 1          & 1          \\
Matches played                 & 0          & 1          & 2          \\
Points tally                 & 0          & 3          & 6          \\
Goal difference              & -        & 2          & 3          \\
Goals scored per match            & 0          & 2          & 2          \\
Goals conceded per match          & 0          & 0          & 0.5        \\
Points per match                   & 0          & 3          & 3          \\
Previous season points tally & 72         & 72         & 72        \\
Previous Season goal difference & 45         & 45         & 45
  \\ \midrule
\multicolumn{4}{l}{Season-specific features} \\ \midrule
  Season                          & 33-34      & 33-34      & 33-34
  \\
Season window                   & August-May & August-May & August-May
  \\ \midrule
\multicolumn{4}{l}{Match-specific features} \\ \midrule
Quarter                         & 3          & 3          & 3         \\
\hline
\end{tabular}
\end{center}
\label{tab-artificial_features}
\end{table*}

\section{Modeling outcomes}
\label{bt}

\subsection{Bradley-Terry models and extensions}
The Bradley-Terry model \citep{Bradley1952} is commonly used to model
paired comparisons, which often arise in competitive sport. For a
binary win/loss outcome, let
\[
y_{ijt} =
\left\{
  \begin{array}{cc}
    1\,, & \text{ if team } i \text{ beats team } j  \text{ at time } t  \\
    0\,, & \text{ if team } j \text{ beats team } i  \text{ at time } t
  \end{array}
\right.
\qquad (i,j =1,\dots,n; \  i\neq j; \  t \in \Re^+) \,,
\]
where $n$ is the number of teams present in the data. The
Bradley-Terry model assumes that
\begin{equation*}
p(y_{ijt}=1) = \frac{\pi_i}{\pi_i+\pi_j} \,,
\end{equation*}
where $\pi_i=\exp (\lambda_i)$, and $\lambda_i$ is understood as the
``strength'' of team $i$. In the original Bradley-Terry formulation,
$\lambda_i$ does not vary with time.

For the purposes of the MLS challenge prediction task, we consider
extensions of the original Bradley-Terry formulation where we allow
$\lambda_i$ to depend on a $p$-vector of time-dependent features
$\bm{x}_{it}$ for team $i$ at time $t$ as
$\lambda_{it} = f(\bm{x}_{it})$ for some function
$f(\cdot)$. Bradley-Terry models can also be equivalently written as
linking the log-odds of a team winning to the difference in strength
of the two teams competing. Some of the extensions below directly
specify that difference.


\subsubsection*{BL: Baseline}

The simplest specification of all assumes that
\begin{equation}
  \label{bl}
  \lambda_{it} = \beta h_{it}\ ,
\end{equation}
where $h_{it} = 1$ if team $i$ is playing at home at time $t$, and
$h_{it} = 0$ otherwise. The only parameter to estimate with this
specification is $\beta$, which can be understood as the difference in
strength when the team plays at home. We use this model to establish a
baseline to improve upon for the prediction task.

\subsubsection*{CS: Constant strengths}

This specification corresponds to the standard Bradley-Terry model
with a home-field advantage, under which
\begin{equation}
  \label{cs}
  \lambda_{it} = \alpha_i + \beta h_{it} \,.
\end{equation}
The above specification involves $n + 1$ parameters, where $n$ is the
number of teams. The parameter $\alpha_i$ represents the
time-invariant strength of the $i$th team.

\subsubsection*{LF: Linear with features}

Suppose now that we are given a vector of features $\bm{x}_{it}$
associated with team $i$ at time $t$. A simple way to model the
team strengths $\lambda_{it}$ is to assume that they are a linear
combination of the features. Hence, in this model we
have
\begin{equation}
  \label{lf}
  \lambda_{it}= \sum_{k=1}^{p} \beta_k x_{itk}\,,
\end{equation}
where $x_{itk}$ is the $k$th element of the feature vector
$\bm{x}_{it}$.

Note that the coefficients in the linear combination are shared
between all teams, and so the number of parameters to estimate is $p$,
where $p$ is the dimension of the feature vector. This specification
is similar to the one implemented in the \texttt{R} package
\texttt{BradleyTerry} \citep{Firth2005}, but without the team specific
random effects.

\subsubsection*{TVC: Time-varying coefficients}

Some of the features we consider, like points tally season (feature 6
in Table~\ref{tab-features}) vary during the season. Ignoring any
special circumstances such as teams being punished, the points
accumulated by a team is a non-decreasing function of the number of
matches the team has played.

It is natural to assume that the contribution of points accumulated to
the strength of a team is different at the beginning of the season
than it is at the end. In order to account for such effects, the
parameters for the corresponding features can be allowed to vary with
the matches played. Specifically, the team strengths can be modeled
as
\begin{equation}
  \label{tvc}
  \lambda_{it}= \sum_{k \in \mathcal{V}} \gamma_k(m_{it}) x_{itk} + \sum_{k \notin \mathcal{V}} \beta_k x_{itk} \,,
\end{equation}
where $m_{it}$ denotes the number of matches that team $i$ has played
within the current season at time $t$ and $\mathcal{V}$ denotes the
set of coefficients that are allowed to vary with the matches played.
The functions $\gamma_k(m_{it})$ can be modeled non-parametrically,
but in the spirit of keeping the complexity low we instead set
$\gamma_k(m_{it}) = \alpha_k + \beta_km_{it}$. With this specification
for $\gamma_k(m_{it})$, TVC is equivalent to LF with the inclusion of
an extra set of features $\{m_{it}x_{itk}\}_{k \in \mathcal{V}}$.

\subsubsection*{AFD: Additive feature differences with time interactions}

For the LF specification, the log-odds of team $i$ beating
team $j$ is
\begin{equation*}
  \lambda_{it} - \lambda_{jt} = \sum_{k=1}^p\beta_k(x_{itk} - x_{jtk})
  \,.
\end{equation*}
Hence, the LF specification assumes that the difference in strength
between the two teams is a linear combination of differences between
the features of the teams. We can relax the assumption of linearity,
and include non-linear time interactions, by instead assuming that
each difference in features contributes to the difference in strengths
through an arbitrary bivariate smooth function $g_k$ that depends on
the feature difference and the number of matches played. We then
arrive at the AFD specification, which can be written as
\begin{equation}
\label{afd}
\lambda_{it} - \lambda_{jt} = \sum_{k \in \mathcal{V}} g_k(x_{itk} - x_{jtk}, m_{it}) + \sum_{k \notin \mathcal{V}} f_k(x_{itk} -
 x_{jtk}) \ , 
\end{equation}
where for simplicity we take the number of matches played to be the number of matches played by the home team.

\subsection{Handling draws}
\label{sec-draws}
The extra outcome of a draw in a soccer match can be accommodated
within the Bradley-Terry formulation in two ways.

The first is to treat win, loss and draw as multinomial ordered
outcomes, in effect assuming that
$\text{``win''} \succ \text{``draw''} \succ \text{``loss''}$, where $\succ$
denotes strong transitive preference. Then, the ordered outcomes can
be modeled using cumulative link models \citep{Agresti2015} with the
various strength specifications. Specifically, let
\[
y_{ijt} =
\begin{cases}
  2\, , & \text{if team } i \text{ beats team } j \text{ at time } t \ ,  \\
  1\, , & \text{if team } i \text{ and } j \text{ draw} \text{ at time } t \ ,\\
  0\, , & \text{if team } j \text{ beats team } i \text{ at time } t \ .

\end{cases}
\]
and assume that $y_{ijt}$ has
\begin{equation}
  \label{ordinal}
p(y_{ijt} \leq y ) = \frac{e^{\delta_{y} + \lambda_{it}}}{e^{\delta_y
    + \lambda_{it}} + e^{\lambda_{jt}}} \, ,
\end{equation}
where $-\infty < \delta_0 \le \delta_1 < \delta_2 = \infty$, and
$\delta_0, \delta_1$ are parameters to be estimated from the
data. \citet{Cattelan2013} and \citet{Kiraly2017} use of this approach
for modeling soccer outcomes.

Another possibility for handling draws is to use the
\citet{Davidson1970} extension of the Bradley-Terry model, under which
\begin{align*}
  p(y_{ijt} = 2 \,|\, y_{ijt} \ne 1) &= \frac{\pi_{it}}{\pi_{it} +
                                       \pi_{jt}}\,, \\
  p(y_{ijt} = 1) &= \frac{\delta\sqrt{\pi_{it}\pi_{jt}}}{\pi_{it} + \pi_{jt} + \delta\sqrt{\pi_{it}\pi_{jt}}}\,, \\
  p(y_{ijt} = 0 \,|\, y_{ijt} \ne 1) &= \frac{\pi_{jt}}{\pi_{it} + \pi_{jt}}\,. \\
\end{align*}
where $\delta$ is a parameter to be estimated from the data.

\subsection{Estimation}

\subsubsection*{Likelihood-based approaches}
The parameters of the Bradley-Terry model extensions presented above
can be estimated by
maximizing the log-likelihood of the multinomial distribution.

The log-likelihood about the parameter vector $\bm{\theta}$ is
\[
  \ell(\bm{\theta}) = \sum_{\{i,j,t\} \in \mathcal{M}}  \sum_{y}
  \mathbb{I}_{[y_{ijt}=y]} \log\Big(p(y_{ijt}=y)\Big)\, ,
\]
where $\mathbb{I_A}$ takes the value $1$ if $A$ holds and $0$ otherwise, and
$\mathcal{M}$ is the set of triplets $\{i,j,t\}$ corresponding to the
matches whose outcomes have been observed.

For estimating the functions involved in the AFD specification, we
represent each $f_k$ using thin plate splines \citep{Wahba1990}, and
enforce smoothness constraints on the estimate of $f_k$ by maximizing
a penalized log-likelihood of the form
\[
\ell^{\text{pen}}(\bm{\theta}) = \ell(\bm{\theta}) - k
\bm{\theta}^TP\bm{\theta}\, ,
\]
where $P$ is a penalty matrix and $k$ is a tuning parameter. For
penalized estimation we only consider ordinal models through the
\texttt{R} package \texttt{mgcv} \citep{Wood2006}, and select $k$ by
optimizing the Generalized Cross Validation criterion
\citep{Golub1979}. Details on the fitting procedure for specifications
like AFD and the implementation of thin plate spline regression in
\texttt{mgcv} can be found in \cite{Wood2003}.


The parameters of the Davidson extensions of the Bradley-Terry model
are estimated by using the BFGS optimization algorithm
\citep{Byrd1995} to minimize $-\ell(\bm{\theta})$.

\subsubsection*{Identifiability}

In the CS model, the team strengths are identifiable only up to an
additive constant, because
$\lambda_i - \lambda_j = (\lambda_i + d) - (\lambda_j + d)$ for any
$d \in \Re$. This unidentifiability can be dealt with by setting the
strength of an arbitrarily chosen team to zero. The CS model was
fitted league-by-league with one identifiability constraint per
league.

The parameters $\delta_0$ and $\delta_1$ in (\ref{ordinal}) are
identifiable only if the specification used for $\lambda_i - \lambda_j$ does not
involve an intercept parameter. An alternative is to include an
intercept parameter in $\lambda_i-\lambda_j$ and fix $\delta_0$ at a value. The
estimated probabilities are invariant to these alternatives, and we
use the latter simply because this is the default in the \texttt{mgcv}
package.

\subsubsection*{Other data-specific considerations}

The parameters in the LF, TVC, and AFD specifications (which involve
features) are shared across the leagues and matches in the data. For
computational efficiency we restrict the fitting procedures to use the
$20, 000$ most recent matches, or less if less is available, at the
time of the first match that a prediction needs to be made. The CS
specification requires estimating the strength parameters
directly. For computational efficiency, we estimate the strength
parameters independently for each league within each country, and only
consider matches that took place in the past calendar year from the
date of the first match that a prediction needs to be made.

\section{Modeling scores} \label{poisson}

\subsection{Model structure}
\label{sec-hpl}
Every league consists of a number of teams $T$, playing against each
other twice in a season (once at home and once away). We indicate the
number of goals scored by the home and the away team in the $g$th
match of the season ($g=1,\ldots,G$) as $y_{g1}$ and $y_{g2}$,
respectively.

The observed goal counts $y_{g1}$ and $y_{g2}$ are assumed to be
realizations of conditionally independent random variables $Y_{g1}$ and $Y_{g2}$,
respectively, with
\begin{eqnarray*}
    Y_{gj} \mid \theta_{gj} \sim \mbox{Poisson}(\theta_{gj}) \ .
\end{eqnarray*}
The parameters $\theta_{g1}$ and $\theta_{g2}$ represent the {\em
  scoring intensity} in the $g$-th match for the home and away team, respectively.

We assume that $\theta_{g1}$ and $\theta_{g2}$ are specified through
the regression structures
\begin{align}
\begin{split}
\eta_{g1} = \log(\theta_{g1}) = \sum_{k=1}^p \beta_k z_{g1k} + \alpha_{h_g} + \xi_{a_g} + \gamma_{h_g,\text{\textit{Sea}}_g} + \delta_{a_g,\text{\textit{Sea}}_g} \ , \\
\eta_{g2} = \log(\theta_{g2}) = \sum_{k=1}^p \beta_k z_{g2k} + \alpha_{a_g} + \xi_{h_g} + \gamma_{a_g,\text{\textit{Sea}}_g} + \delta_{h_g,\text{\textit{Sea}}_g} \ .
\end{split}
\label{linpred}
\end{align}
The indices $h_g$ and $a_g$ determine the home and away team for match
$g$ respectively, with $h_g, a_g \in \{1, \ldots, T\}$. The parameters
$\beta_1, \ldots, \beta_p$ represent the effects corresponding to the
observed match- and team-specific features $z_{gj1},\ldots, z_{gjp}$,
respectively, collected in a $G \times 2p$ matrix $\bm{Z}$. The other
effects in the linear predictor $\eta_{gj}$ reflect assumptions of
exchangeability across the teams involved in the
matches. Specifically, $\alpha_t$ and $\xi_t$ represent the latent
attacking and defensive ability of team $t$ and are assumed to be
distributed as
\[
  \alpha_t \mid \sigma_\alpha \sim\mbox{Normal}(0,\sigma_\alpha^2) \qquad \mbox{and} \qquad
  \xi_t \mid \sigma_\xi \sim \mbox{Normal}(0,\sigma_\xi^2).
\]
We used vague log-Gamma priors on the precision parameters
$\tau_\alpha=1/\sigma^2_\alpha$ and $\tau_\xi=1/\sigma^2_\xi$. In
order to account for the time dynamics across the different seasons,
we also include the latent interactions $\gamma_{ts}$ and
$\delta_{ts}$ between the team-specific attacking and defensive
strengths and the season $s \in \{ 1, \ldots, S \}$, which were
modeled using autoregressive specifications with
\[
  \gamma_{t1} \mid \sigma_\varepsilon, \rho_\gamma
  \sim\mbox{Normal}\left(0,\sigma^2_\varepsilon(1-\rho_\gamma^2)\right),
  \quad \gamma_{ts}=\rho_\gamma\gamma_{t,s-1}+\varepsilon_{s}, \quad
  \varepsilon_s \mid  \sigma_\varepsilon
  \sim\mbox{Normal}(0,\sigma_\varepsilon^2) \quad (s = 2, \ldots, S)\,,
\]
and
\[\delta_{t1} \mid  \sigma_\varepsilon, \rho_\delta
  \sim\mbox{Normal}\left(0,\sigma^2_\varepsilon(1-\rho_\delta^2)\right),
  \quad \delta_{ts}=\rho_\delta\delta_{t,s-1}+\varepsilon_{s}, \quad
  \varepsilon_s \mid  \sigma_\varepsilon
  \sim\mbox{Normal}(0,\sigma_\varepsilon^2) \quad (s = 2, \ldots, S) \, .
\]
For the specification of prior distributions for the hyperparameters
$\rho_\gamma, \rho_\delta, \sigma_\epsilon$ we used the default
settings of the \texttt{R-INLA} package \citep[version
17.6.20]{Lindgren2015}, which we also use to fit the model (see
Subsection \ref{estpoisson}). Specifically, \texttt{R-INLA} sets vague
Normal priors (centred at $0$ with large variance) on suitable
transformations (e.g. log) of the hyperparameters with unbounded
range.

\subsection{Estimation} \label{estpoisson}

The hierarchical Poisson log-linear model (HPL) of
Subsection~\ref{sec-hpl} was fitted using INLA
\citep{Rue2009}. Specifically, INLA avoids time-consuming MCMC
simulations by numerically approximating the posterior densities for
the parameters of latent Gaussian models, which constitute a wide
class of hierarchical models of the form
\begin{eqnarray*}
Y_i \mid \bm{\phi},\bm\psi & \sim & p(y_i\mid \bm\phi,\bm\psi) \ , \\
\bm\phi \mid \bm\psi & \sim & \mbox{Normal}\left(\bm{0},\bm{Q}^{-1}(\bm\psi)\right) \ , \\
\bm\psi & \sim & p(\bm{\psi}) \ ,
\end{eqnarray*}
where $Y_i$ is the random variable corresponding to the observed
response $y_i$, $\bm\phi$ is a set of parameters (which may have a
large dimension) and $\bm\psi$ is a set of hyperparameters.


The basic principle is to approximate the posterior densities for
$\bm\psi$ and $\bm\phi$ using a series of nested Normal
approximations. The algorithm uses numerical optimization to find the
mode of the posterior, while the marginal posterior distributions are
computed using numerical integration over the hyperparameters. The
posterior densities for the parameters of the HPL model are computed
on the available data for each league.

To predict the outcome of a future match, we simulated $1000$ samples
from the joint approximated predictive distribution of the number of
goals $\tilde{Y}_1$, $\tilde{Y}_{2}$, scored in the future match by
the home and away teams respectively, given features
$\tilde{\bm{z}}_{j} = (\tilde{z}_{j1}, \ldots,
\tilde{z}_{j2})^\top$. Sampling was done using the
\texttt{inla.posterior.sample} method of the \texttt{R-INLA}
package. The predictive distribution has a probability mass function
of the form
\[
  p\left(\tilde{y}_{1}, \tilde{y}_{2} \mid \bm{y}_1, \bm{y}_2,
    \tilde{\bm{z}}_1, \tilde{\bm{z}}_2, \bm{Z} \right) = \int
  p\left(\tilde{y}_{1},\tilde{y}_{2} \mid \bm{\nu},
    \tilde{\bm{z}}_1, \tilde{\bm{z}}_2 \right) p\left(\bm{\nu} \mid
    \bm{y}_1, \bm{y}_2, \bm{Z}\right)d\bm{\nu} \, ,
\]
where the vector $\bm{\nu}$ collects all model parameters. We then
compute the relative frequencies of the events
$\tilde{Y}_{1}> \tilde{Y}_{2}$, $\tilde{Y}_{1}=\tilde{Y}_{2}$, and
$\tilde{Y}_{1}<\tilde{Y}_{2}$, which correspond to home win, draw, and
loss respectively.

\section{Validation framework}
\label{validationsec}

\subsection{MLS challenge}

The MLS challenge consists of predicting the outcomes (win, draw,
loss) of 206 soccer matches from 52 leagues that take place between
31st March 2017 and 10th April 2017. The prediction performance of
each submission was assessed in terms of the average ranked
probability score (see Subsection~\ref{sec-criteria}) over those
matches. To predict the outcomes of these matches, the challenge
participants have access to over 200,000 matches up to and including the 21st March 2017,
which can be used to train a classifier. 

\begin{figure}[t]
\centering
\includegraphics[scale=0.7]{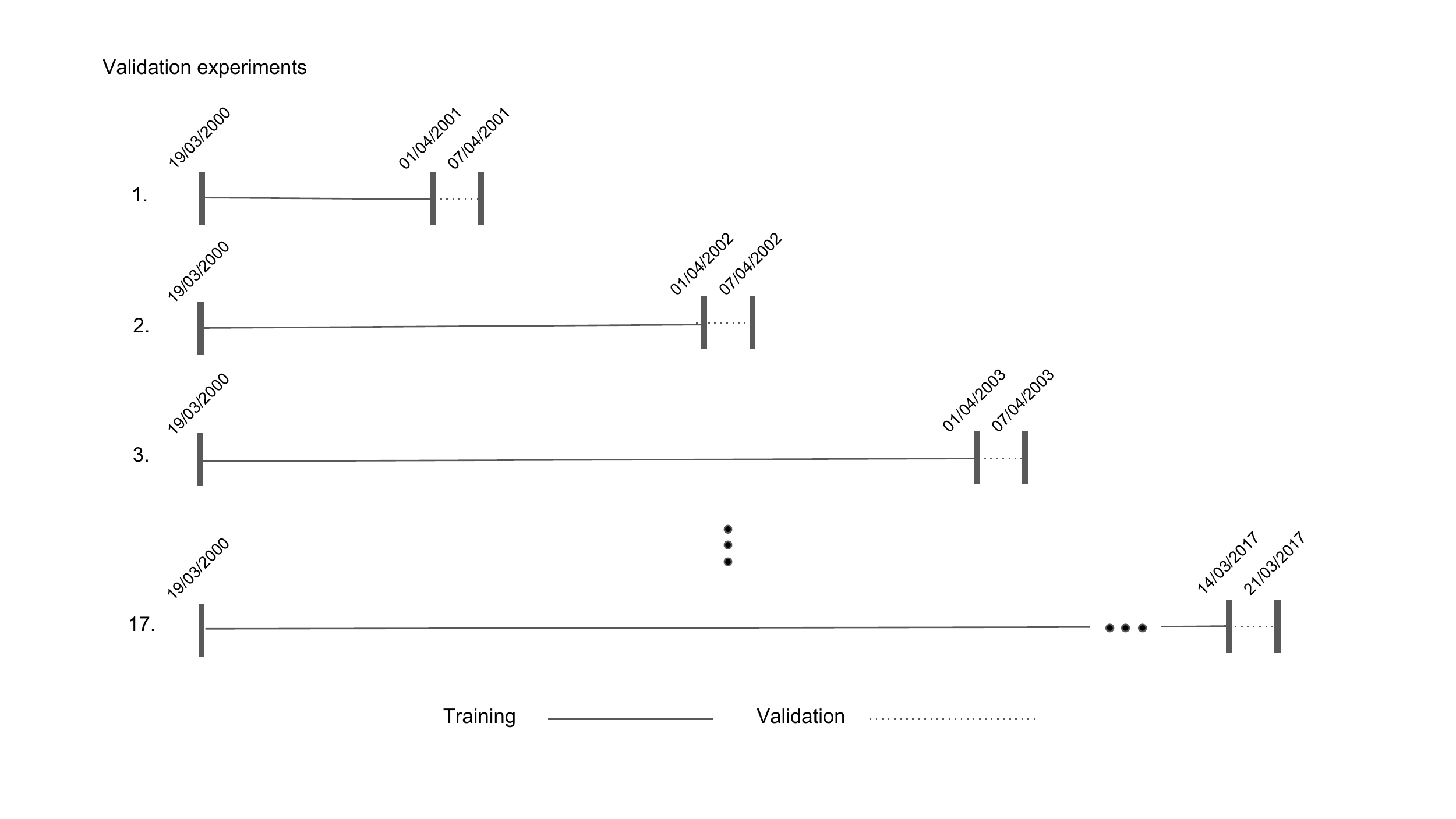}
\caption{The sequence of experiments that constitute the validation
  framework, visualizing their corresponding training and prediction
  periods. \label{validation}}
\end{figure}

In order to guide the choice of the model that is best suited to make
the final predictions, we designed a validation framework that
emulates the requirements of the MLS Challenge. We evaluated the
models in terms of the quality of future predictions, i.e.~predictions
about matches that happen after the matches used for training. In
particular, we estimated the model parameters using data from the
period before 1st April of each available calendar year in the data,
and examined the quality of predictions in the period between 1st and
7th April of that year. For 2017, we estimated the model parameters
using data from the period before 14th March 2017, and examined the
quality of predictions in the period between 14th and 21st March 2017. Figure~\ref{validation} is a pictorial representation of the validation framework, illustrating the sequence of experiments and
the duration of their corresponding training and validation periods.

\subsection{Validation criteria}
\label{sec-criteria}
The main predictive criterion we used in the validation framework is
the ranked probability score, which is also the criterion that was
used to determine the outcome of the challenge. Classification
accuracy was also computed.

\subsubsection*{Ranked probability score}

Let $R$ be the number of possible outcomes (e.g.~$R = 3$ in soccer)
and $\bm{p}$ be the $R$-vector of predicted probabilities with $j$-th
component $p_j \in [0,1]$ and $p_1 + \ldots + p_R = 1$. Suppose that
the observed outcomes are encoded in an $R$-vector $\bm{a}$ with $j$-th
component $a_j \in \{0,1\}$ and $a_1 + \ldots + a_r = 1$. The ranked
probability score is defined as\
\begin{equation}
    {\rm RPS} = \frac{1}{r-1} {\sum_{i =
        1}^{r-1}\left\{\sum_{j=1}^i\left(p_j - a_j\right)\right\}^2}
    \, .
\end{equation}
The ranked probability score was introduced by \citet{Epstein1969}
\citep[see also,][for a general review of scoring rules]{Gneiting2007}
and is a strictly proper probabilistic scoring rule, in the sense that
the true odds minimize its expected value \citep{Murphy1969}.

\subsubsection*{Classification accuracy}
Classification accuracy measures how often the classifier makes the
correct prediction, i.e. how many times the outcome with the
maximum estimated probability of occurence actually occurs.
\begin{center}
  \begin{table*}[ht]
    \caption{Illustration of the calculation of the ranked probability
      score and classification accuracy on artificial data.}
    {
\hfill{}
\footnotesize 
\begin{tabular}{lllllllllll}
\hline
\multicolumn{3}{l}{Observed outcome} & \multicolumn{3}{l}{Predicted probabilities} & \multicolumn{3}{l}{Predicted outcome} & \multirow{2}{*}{RPS} & \multirow{2}{*}{Accuracy} \\
$a_1$       & $a_2$      & $a_3$      & $p_1$       & $p_2$       & $p_3$   & $o_1$       & $o_2$       & $o_3$  & &   \\
\hline
1       & 0      & 0      & 1        & 0        & 0       & 1        & 0        & 0  & 0        & 1              \\
1       & 0      & 0      & 0        & 1        & 0       & 0        & 1        & 0  & 0.5      & 0              \\
1       & 0      & 0      & 0        & 0        & 1       & 0        & 0        & 1  & 1        & 0              \\
1       & 0      & 0      & 0.8      & 0.2      & 0       & 1        & 0        & 0  & 0.02     & 1              \\
0       & 1      & 0      & 0.33     & 0.33     & 0.34    & 0        & 0        & 1  & 0.11     & 0              \\
\hline
\end{tabular}}
\hfill{}
\label{rpsexample}
\end{table*}
\end{center}

Table~\ref{rpsexample} illustrates the calculations leading to the
ranked probability score and classification accuracy for several
combinations of $\bm{p}$ and $\bm{a}$. The left-most group of three
columns gives the observed outcomes, the next group gives the
predicted outcome probabilities, and the third gives the predicted
outcomes using maximum probability allocation. The two columns in the
right give the ranked probability scores and classification
accuracies. As shown, a ranked probability score of zero
indicates a perfect prediction (minimum error) and a ranked
probability score of one indicates a completely wrong prediction
(maximum error).

The ranked probability score and classification accuracy for a
particular experiment in the validation framework are computed by
averaging over their respective values over the matches in the
prediction set. The uncertainty in the estimates from each experiment
is quantified using leave-one-match out jackknife \citep{efron1982},
as detailed in step~9 of Algorithm~\ref{validationalgo}.

\subsection{Meta-analysis}

The proposed validation framework consists of $K = 17$ experiments,
one for each calendar year in the data. Each experiment results in
pairs of observations $(s_i, \hat{\sigma_i}^2)$, where $s_i$ is the
ranked probability score or classification accuracy from the $i$th
experiment, and $\hat{\sigma_i}^2$ is the associated jackknife
estimate of its variance $(i = 1, \ldots , K)$.

We synthesized the results of the experiments using meta-analysis
\citep{DerSimonian1986}. Specifically, we make the working assumptions
that the summary variances $\hat{\sigma_i}^2$ are estimated
well-enough to be considered as known, and that $s_1, \ldots, s_K$ are
realizations of random variables $S_1, \ldots, S_K$, respectively,
which are independent conditionally on independent random effects
$U_1, . . . , U_K$, with
\[
    {S_i \mid U_i} \sim \mbox{Normal} (\alpha + U_i, \hat{\sigma_i}^2) \, ,
\]
and
\[
    U_i \sim \mbox{Normal} (0, \tau^2) \, .
\]

The parameter $\alpha$ is understood here as the overall ranked
probability score or classification accuracy, after accounting for the
heterogeneity between the experiments.

The maximum likelihood estimate of the overall ranked probability or
classification accuracy is then the weighted average
\[
    \hat{\alpha} = \frac{\sum w_i s_i}{\sum w_i} \ ,
\]
where $w_i = (\hat{\sigma_i}^2 + \hat{\tau}^2)^{-1}$ and
$\hat{\tau}^2$ is the maximum likelihood estimate of $\tau^2$. The
estimated standard error for the estimator of the overall score
$\hat\alpha$ can be computed using the square root of the inverse
Fisher information about $\alpha$, which ends up being
$(\sum_{i = 1}^K w_i)^{-1/2}$.

The assumptions of the random-effects meta-analysis model
(independence, normality and fixed variances) are all subject to
direct criticism for the validation framework depicted in
Figure~\ref{validation} and the criteria we consider; for example, the
training and validation sets defined in the sequence of experiments in
Figure~\ref{validation} are overlapping and ordered in time, so the
summaries resulting from the experiment are generally correlated. We
proceed under the assumption that these departures are not severe
enough to influence inference and conclusions about $\alpha$.

\subsection{Implementation}

Algorithm~\ref{validationalgo} is an implementation of the validation
framework in pseudo-code. Each model is expected to have a training
method which trains the model on data, and a prediction method which
returns predicted outcome probabilities for the prediction set. We
refer to these methods as {\tt train} and {\tt predict} in the
pseudo-code.

\begin{algorithm} [t!]
  \caption{Pseudo-code for the validation framework} \label{validationalgo}
  \footnotesize
  \begin{algorithmic}[1]
    \Require{
    \Statex $\bm{x}_1, \ldots , \bm{x}_G$ \Comment{feature vectors for all $G$}
    matches in the data set
    \Statex $d_1 \leq \ldots \leq d_G$ \Comment{$d_g$ is the match date of match $g \in \{1,\ldots,G\}$}
    \Statex $\bm{o}_1, \ldots , \bm{o}_G$ \Comment{match outcomes}
    \Statex train: $\{ \bm{x}_g, \bm{o}_g :  g \in A\} \to f(\cdot) $
    \Comment{Training algorithm}
    \Statex predict: $\{ \bm{x}_g : g \in B\}, f(\cdot) \to
    \{\bar{\bm{o}}_g: g \in B\}$    \Comment{Prediction algorithm}
    \Statex criterion: $\{ \bm{o}_g, \bar{\bm{o}}_g : g \in B \} \to
    \{v_g: g \in B\}$ \Comment{observation-wise criterion values}
    \Statex $D_1, \ldots, D_{T}$ \Comment{Cut-off dates for training
      for experiments}
    \Statex meta-analysis: $\{ s_i, \hat\sigma_i^2 : i \in \{1,
      \ldots, T\}\} \to \hat\alpha $
    \Comment{Meta-analysis algorithm}

    }
    \Statex
    \Ensure{
      $\hat\alpha$ \Comment{Overall validation metric}
    }
    \Statex
    \For{$i \gets 1 \textrm{ to } T$}

    \Let{$A$}{$\{g: d_g \le D_t\}$}
    \Let{$B$}{$\{g: D_t < d_g \le D_t + 10\text{days}\}$}

    \Let{$n_B$}{${\rm dim}(B)$}

    \Let{$f(\cdot)$}{train($\{ \bm{x}_g, \bm{o_g} : g \in A \}$)} \Comment{fit the model}
    \Let{$\{ \bar{\bm{o}}_g: g\in B\}$}{predict($\{\bm{x}_g: g \in B\}, f(\cdot)$)} \Comment{get predictions}

    \Let{$\{v_g: g \in B\}$}{criterion$(\{\bm{o}_g, \bar{\bm{o}}_g: g \in
      B\})$}

    \Let{$s_i$}{$\frac{1}{n_B} \sum_{g \in B} v_g$}

    \Let{$\hat\sigma_i^2$}{$\frac{n_B}{n_B - 1}\sum_{g \in B} \left(
        \frac{\sum_{h\in B/\{g\}} v_h}{n_B - 1} - s_i \right)^2$}

    \EndFor
    \Statex
    \Let{$\hat\alpha$}{meta-analysis($\{s_i, \hat\sigma_i^2: i \in \{1, \ldots, T\}$)}
  \end{algorithmic}
\end{algorithm}
\section{Results}
\label{results}
In this section we compare the predictive performance of the various
models we implemented as measured by the validation framework
described in Section \ref{validationsec}. Table~\ref{modref} gives the
details of each model in terms of features used, the handling of draws
(ordinal and Davidson, as in Subsection~\ref{sec-draws}), the
distribution whose parameters are modeled, and the estimation
procedure that has been used.

The sets of features that were used in the LF, TVC, AFD and HPL
specifications in Table~\ref{modref} resulted from ad-hoc
experimentation with different combinations of features in the LF
specification. All instances of feature 13 refer to the least squares
ordinal rank (see Subsection 2.5 of the supplementary material). The
features used in the HPL specification in (\ref{linpred}) have been
chosen prior to fitting to be home and newly promoted (features 1 and
2 in Table~\ref{tab-features}),
the difference in form and points tally (features 4 and 6 in
Table~\ref{tab-features}) between the two teams competing in match
$g$, and season and quarter (features 15 and 16 in
Table~\ref{tab-features}) for the season that match $g$ takes place.



\begin{table*}[ht]
  \caption{Description of each model in Section~\ref{bt} and
    Section~\ref{poisson} in terms of features used, the handling of
    draws, the distribution whose parameters are modeled, and the
    estimation procedure that was used. The suffix $(t)$ indicates
    features with coefficients varying with matches played (feature 5
    in Table~\ref{tab-features}). The model indicated by $\dagger$ is
    the one we used to compute the probabilities for the submission
    to the MLS challenge. The acronyms are as follows:   BL: Baseline
    (home advantage); CS: Bradley-Terry with constant strengths; LF:
    Bradley-Terry with linear features; TVC:  Bradley-Terry with
    time-varying coefficients; AFD: Bradley-Terry with additive
    feature differences and time interactions; HPL:Hierarchical
    Poisson log-linear model.} \vspace{0.2cm}
  \centering
	\footnotesize
\begin{tabular}{rrrlr}
  \hline
  Model & Draws & Features &  Distribution & Estimation \\
  \hline
  BL (\ref{bl}) & Davidson & 1 &  Multinomial & ML  \\
  BL (\ref{bl}) & Ordinal & 1 &  Multinomial & ML  \\
  CS  (\ref{cs}) & Davidson & 1  & Multinomial & ML  \\
  CS  (\ref{cs}) & Ordinal & 1 &  Multinomial & ML  \\
  LF  (\ref{lf}) & Davidson & 1, 6, 7, 12, 13 &  Multinomial  & ML  \\
  LF (\ref{lf}) & Ordinal & 1, 6, 7, 12, 13 & Multinomial & ML  \\
  TVC (\ref{tvc}) & Davidson & 1, 6$(t)$, 7$(t)$, 12$(t)$, 13 & Multinomial  & ML  \\
  TVC (\ref{tvc}) & Ordinal & 1, 6$(t)$, 7$(t)$, 12$(t)$, 13 & Multinomial  & ML  \\
  AFD (\ref{afd}) & Davidson & 1, 6$(t)$, 7$(t)$, 12$(t)$, 13 & Multinomial & MPL  \\
  HPL (\ref{linpred}) & & 1, 2, 4, 6, 15, 16 & Poisson & INLA  \\
  ($\dagger$) TVC (\ref{tvc}) & Ordinal & 1, 2, 3, 4, 6$(t)$, 7$(t)$, 11$(t)$ & Multinomial  & ML  \\
  \hline
\end{tabular}
\label{modref}
\end{table*}

For each of the models in Table~\ref{modref}, Table~\ref{resultstab}
presents the ranked probability score and classification accuracy as
estimated from the validation framework in
Algorithm~\ref{validationalgo}, and as calculated for the matches in
the test set for the challenge.

The results in Table~\ref{resultstab} are indicative of the good
properties of the validation framework of Section~\ref{validationsec}
in accurately estimating the performance of the classifier on unseen
data. Specifically, and excluding the baseline model, the sample
correlation between overall ranked probability score and the average
ranked probability score from the matches on the test set is
$0.973$. The classification accuracy seems to be underestimated by the
validation framework.

\begin{figure}[t]
	\vspace{-0.5cm}
	\centering
	\includegraphics[scale=0.6]{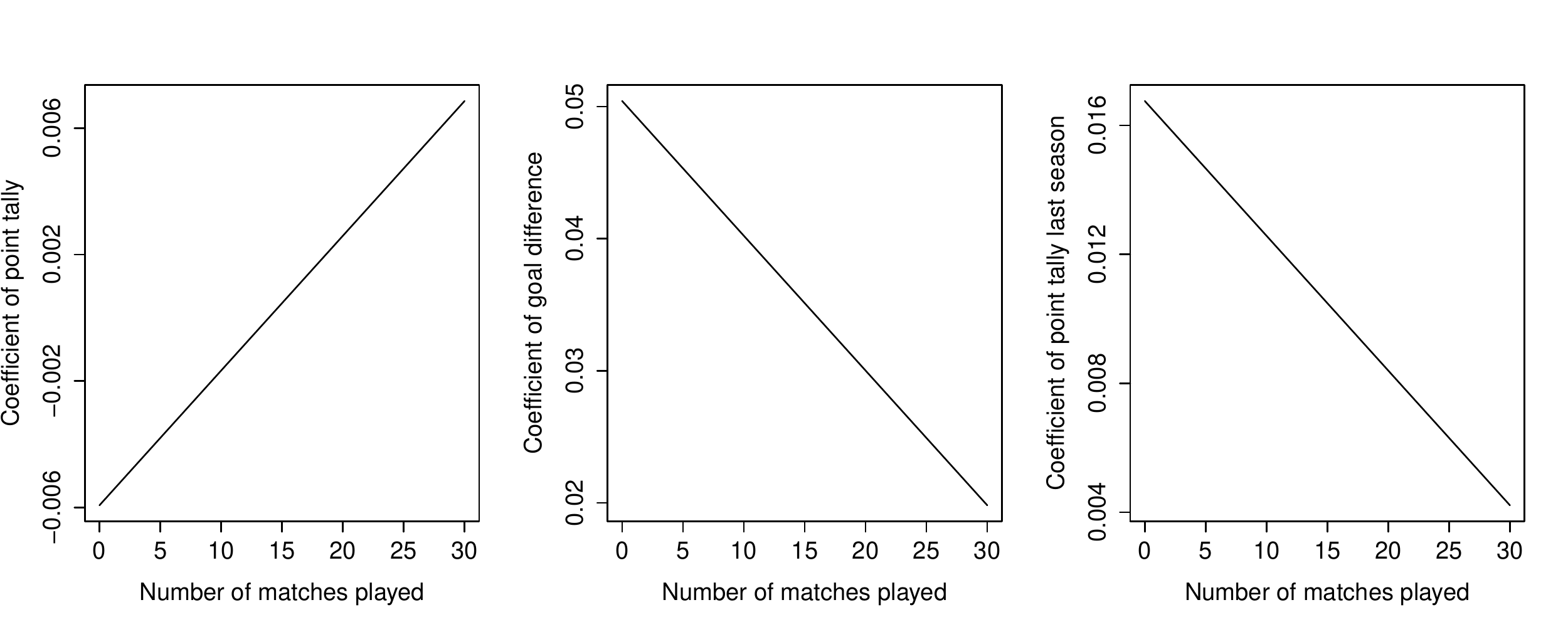}
	\caption{Plots of the time-varying coefficients in the TVC
          model that is indicated by $\dagger$ in
          Table~\ref{resultstab}, which is the model we used to
          compute the probabilities for our submission to the MLS
          challenge. 
}
              \label{coefplots}
\end{figure}

The TVC model that is indicated by $\dagger$ in Table~\ref{resultstab}
is the model we used to compute the probabilities for our submission
to the MLS challenge. Figure \ref{coefplots} shows the estimated
time-varying coefficients for the TVC model. The remaining parameter
estimates are $0.0410$ for the coefficient of form, $-0.0001$ for the
coefficient of days since previous match, and $0.0386$ for the
coefficient of newly promoted. Of all the features included, only goal
difference and point tally last season had coefficients for which we
found evidence of difference from zero when accounting for all other
parameters in the model (the $p$-values from individual Wald tests are
both less than $0.001$).

After the completion of the MLS challenge we explored the potential of
new models and achieved even smaller ranked probability scores than
the one obtained from the TVC model. In particular, the best
performing model is the HPL model in Subsection~\ref{sec-hpl} (starred
in Table~\ref{resultstab}), followed by the AFD model which achieves a
marginally worse ranked probability score. It should be noted here
that the LF models are two simpler models that achieve performance
that is close to that of HPL and AFD, without the inclusion of random
effects, time-varying coefficients, or any non-parametric
specifications.

The direct comparison between the ordinal and Davidson extensions of
Bradley-Terry type models indicates that the differences tend to be
small, with the Davidson extensions appearing to perform better.

We also tested the performance of HPL in terms of predicting actual scores of matches using the validation framework, comparing to a baseline method that always predicts the average goals scored by home and away teams respectively in the training data it receives. Using root mean square error as an evaluation metric, HPL achieved a score of 1.0011 with estimated standard error 0.0077 compared to the baseline which achieved a score of 1.0331 with estimated standard error 0.0083.
\begin{table}[ht]
  \caption{Ranked probability score and classification accuracy for
    the models in Table~\ref{modref}, as estimated from the validation
    framework of Section~\ref{validationsec} (standard errors are in
    parentheses) and from the matches in the test set of the
    challenge. The model indicated by $\dagger$ is the one we used to
    compute the probabilities for the submission to the MLS challenge,
    while the one indicated by $*$ is the one that achieves the lowest
    estimated ranked probability score.}
\label{resultstab}
\centering
\footnotesize 
\begin{tabular}{rrrrrrrrr}
  \toprule
          &   & & \multicolumn{3}{c}{Ranked probability score} &
                                                                 \multicolumn{3}{c}{Accuracy}
  \\ \cmidrule{3-8}
            & Model & Draws & \multicolumn{2}{c}{Validation} & Test &
                                              \multicolumn{2}{c}{Validation}
    & Test \\ \midrule
            & BL  & Davidson & 0.2242 & (0.0024) & 0.2261 & 0.4472 & (0.0067) & 0.4515 \\
            & BL  & Ordinal & 0.2242 & (0.0024) & 0.2261 & 0.4472 & (0.0067) & 0.4515 \\
            & CS  & Davidson         & 0.2112 & (0.0028) & 0.2128 & 0.4829 & (0.0073) & 0.5194 \\
            & CS  & Ordinal          & 0.2114 & (0.0028) & 0.2129 & 0.4779 & (0.0074) & 0.4951 \\
            & LF  & Davidson         & 0.2088 & (0.0026) & 0.2080 & 0.4849 & (0.0068) & 0.5049 \\
            & LF  & Ordinal          & 0.2088 & (0.0026) & 0.2084 & 0.4847 & (0.0068) & 0.5146 \\
            & TVC & Davidson         & 0.2081 & (0.0026) & 0.2080 & 0.4898 & (0.0068) & 0.5049 \\
            & TVC & Ordinal          & 0.2083 & (0.0025) & 0.2080 & 0.4860 & (0.0068) & 0.5097 \\
            & AFD & Ordinal          & 0.2079 & (0.0026) & 0.2061 & 0.4837 & (0.0068) & 0.5194 \\
  $\star$   & HPL &                  & 0.2073 & (0.0025) & 0.2047 & 0.4832 & (0.0067) & 0.5485 \\
  $\dagger$ & TVC & Ordinal          & 0.2085 & (0.0025) & 0.2087 & 0.4865 & (0.0068) & 0.5388 \\
  \bottomrule
\end{tabular}
\end{table}

\section{Conclusions and discussion}
\label{conclusion}

We compared the performance of various extensions of Bradley-Terry
models and a hierarchical log-linear Poisson model for the prediction
of outcomes of soccer matches. The best performing Bradley-Terry model
and the hierachical log-linear Poisson model delivered similar
performance, with the hierachical log-linear Poisson model doing
marginally better. 


Amongst the Bradley-Terry specifications, the best performing one is
AFD, which models strength differences through a semi-parametric
specification involving general smooth bivariate functions of features
and season time. Similar but lower predictive performance was achieved
by the Bradley-Terry specification that models team strength in terms
of linear functions of season time. Overall, the inclusion of features
delivered better predictive performance than the simpler Bradley-Terry
specifications. In effect, information is gained by relaxing the
assumption that each team has constant strength over the season and
across feature values. The fact that the models with time varying
components performed best within the Bradley-Terry class of models
indicates that enriching models with time-varying specifications can
deliver substantial improvements in the prediction of soccer outcomes.

All models considered in this paper have been evaluated using a novel,
context-specific validation framework that accounts for the temporal
dimension in the data and tests the methods under gradually increasing
information for the training. The resulting
experiments are then pooled together using meta-analysis in order to
account for the differences in the uncertainty of the validation
criterion values by weighing them accordingly.

The meta analysis model we employed operates under the working
assumption of independence between the estimated validation criterion
values from each experiment. This is at best a crude assumption in
cases like the above where data for training may be shared between
experiments. Furthermore, the validation framework was designed to
explicitly estimate the performance of each method only for a
pre-specified window of time in each league, which we have set close
to the window where the MLS challenge submissions were being
evaluated. As a result, the conclusions we present are not
generalizable beyond the specific time window that was
considered. Despite these shortcomings, the results in
Table~\ref{resultstab} show that the validation framework delivered
accurate estimates of the actual predictive performance of each
method, as the estimated average predictive performances and the actual performances on the test set (containing matches between 31st March and 10th April, 2017) were very close.

The main focus of this paper is to provide a workflow for predicting
soccer outcomes, and to propose various alternative models for the
task. Additional feature engineering and selection, and alternative
fitting strategies can potentially increase performance and are worth
pursuing. For example, ensemble methods aimed at improving predictive
accuracy like calibration, boosting, bagging, or model averaging
\cite[for an overview, see][]{Dietterich2000} could be utilized to
boost the performance of the classifiers that were trained in this
paper.

A challenging aspect of modeling soccer outcomes is devising ways to
borrow information across different leagues. The two best performing
models (HPL and AFD) are extremes in this respect; HPL is trained on
each league separately while AFD is trained on all leagues
simultaneously, ignoring the league that teams belong to. Further
improvements in predictive quality can potentially be achieved by using a hierarchical model that takes into account which league teams belong to but also allows for sharing of information between leagues. 



\section{Supplementary material}
The supplementary material document contains two sections. Section 1
provides plots of the number of matches per number of goals scored by
the home and away teams, by country, for a variety of arbitrarily
chosen countries. These plots provide evidence of a home
advantage. Section 2 details approaches for obtaining team rankings
(feature 13 in Table~\ref{tab-artificial_features}) based on the
outcomes of the matches they played so far.

\section{Acknowledgements and authors' contributions}

This work was supported by The Alan Turing Institute under the EPSRC
grant EP/N510129/1.

The authors are grateful to Petros Dellaportas, David Firth,
Istv\'{a}n Papp, Ricardo Silva, and Zhaozhi Qian for helpful
discussions during the challenge. Alkeos Tsokos, Santhosh Narayanan
and Ioannis Kosmidis have defined the various Bradley-Terry
specifications, and devised and implemented the corresponding
estimation procedures and the validation framework. Gianluca Baio
developed the hierarchical Poisson log-linear model and the associated
posterior inference procedures. Mihai Cucuringu did extensive work on
feature extraction using ranking algorithms. Gavin Whitaker carried
out core data wrangling tasks and, along with Franz Kir\'aly, worked
on the initial data exploration and helped with the design of the
estimation-prediction pipeline for the validation experiments. Franz
Kir\'aly also contributed to the organisation of the team meetings and
communication during the challenge. All authors have discussed and
provided feedback on all aspects of the challenge, manuscript
preparation and relevant data analyses.



\clearpage



\section*{Supplementary material}

# Supplementary material for "Modeling outcomes of soccer matches"

Alkeos Tsokos[1], Santhosh Narayanan[2], Ioannis Kosmidis[2,3], Gianluca Baio[1], Mihai Cucuringu[3,4], Gavin Whitaker[1], and Franz Király[1,3]

[1]University College London
[2]University of Warwick
[3]The Alan Turing Institute
[4]University of Oxford

August 3, 2018

## 1  Home advantage

Figure 1 displays plots of the number of matches per number of goals scored by the home (dark grey) and away teams (light grey), by country, for a variety of arbitrarily chosen countries. The plots demonstrate that the home advantage that appears in terms of goals scored when looking at the data for all countries as a whole, holds when looking at individual countries as well.

## 2  Extracting features via ranking from noisy pairwise comparisons

This section details our approach to deriving features based on the ranking of teams obtained from aggregated historical matches between pairs of teams. The main purpose of this section is to provide a high level overview of a number of algorithms for ranking from pairwise comparison data, including several state-of-the-art approaches for this task, which are summarized in Table 1. Broadly speaking, there are two choices to make: one is the particular ranking algorithm used, and the other is the input matrix whose individual entries serve as a proxy for the rank offset between pairs of teams. In our numerical experiments, we chose to consider five different algorithms, and two different types of input matrices, detailed in the next paragraph. These choices are by no means exhaustive, especially in the latter direction, however they serve as a good basis towards exploring this approach.

We summarize in Table 2 the numerical results obtained from each of the algorithms considered in this section, when the resulting features were used on their own in the LF Bradley-Terry formulation (see Section 3.1 in the main text). All algorithms take as input a skew-symmetric matrix $M$, whose entries are interpreted as a proxy for the pairwise rank offset between pairs of teams. To each ranking algorithm, we append the suffix "-card" when the entries in the pairwise comparison matrix $M$ are based on the average goal differential computed over the considered historical window, which in our case was chosen to be the previous three seasons and the current season (if a pair of teams did play a game this season), using equal weights. For example, if a pair of teams played a total of four matches in the previous three seasons, and one match in the current season thus far, then $M_{ij}^{card}$ holds the aggregated goal differential over the five matches.

1

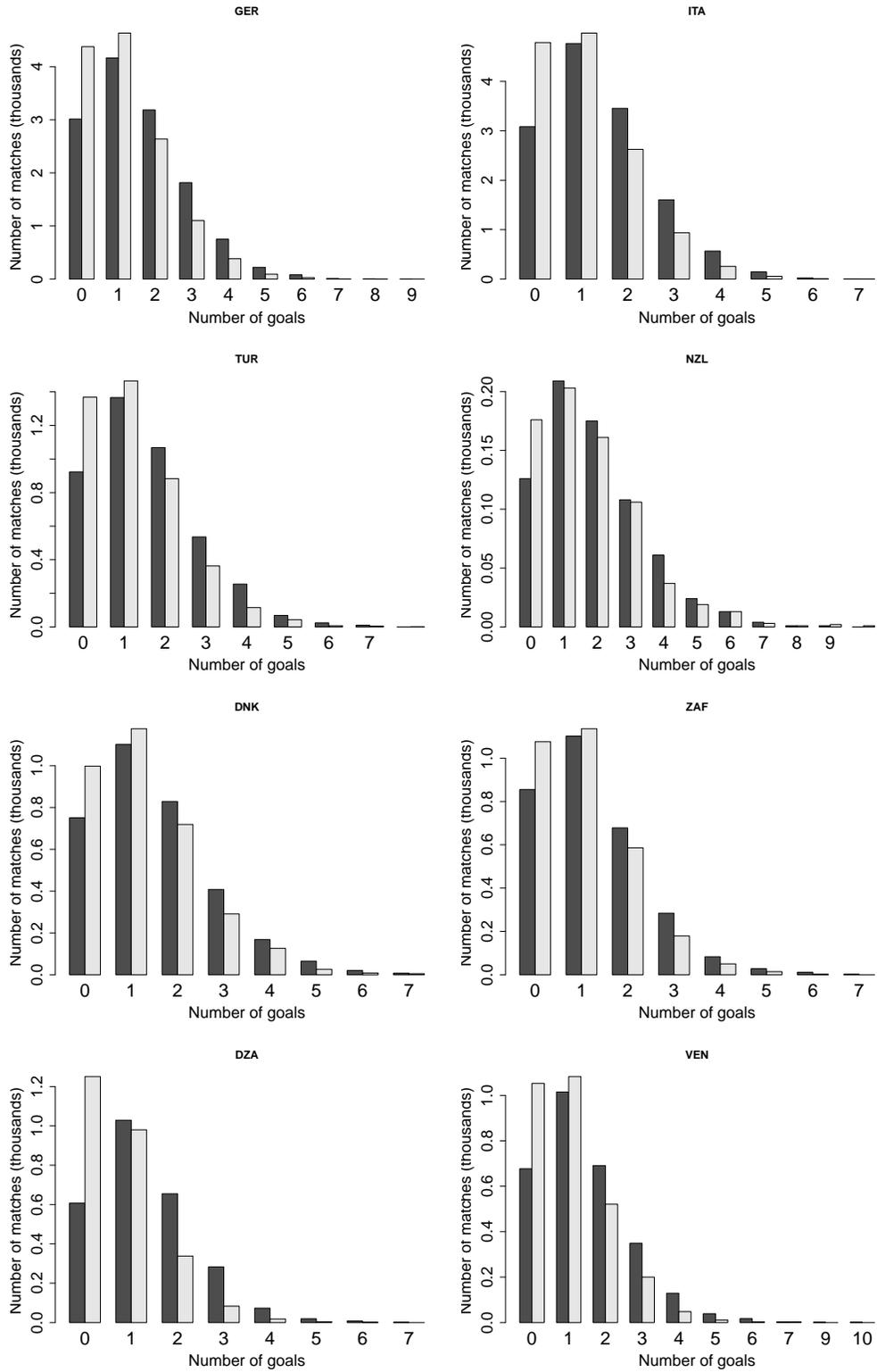

Figure 1: The number of matches per number of goals scored by the home (dark grey) and away teams (light grey), by country, for a variety of arbitrarily chosen countries.

2

We append the suffix "*-ord*" if only ordinal information is available, i.e. if the input to the ranking algorithms is the matrix $M_{ij}^{ord} = \text{sign}(M_{ij}^{ord})$, that captures, for a pair of teams, which team scored more goals on aggregate over the previous direct matches. We denoted the resulting time dependent comparison matrices by $M^{card}$ and $M^{ord}$, respectively, and remark that there are numerous other options for building such matrices. For example, instead of relying on the goal differentials over the previous three seasons and the current season, one could

- pool data only from the previous season and the current one,
- aggregate historical data from the past $k$ seasons, where the weights of the matches decay harmonically with time,
- take into account only counts of the number of wins and losses, as opposed to the actual goal differentials.

Due to time considerations, we have not explored all of these possibilities in our simulations.

We build on the recent work of Cucuringu (2016), which considers the classical problem of establishing a statistical ranking of a set of $n$ items given a set of inconsistent and incomplete pairwise comparisons between such items. Instantiations of this problem occur in numerous applications in data analysis, including analysis of sports data. We formulate the above problem of ranking with incomplete noisy information as an instance of the *group synchronization* problem over the group SO(2) of planar rotations, whose usefulness has been demonstrated in numerous applications in recent years in areas such as computer vision and graphics, sensor network localization and structural biology. Its least squares solution can be approximated by either a spectral or a semidefinite programming (SDP) relaxation, followed by a rounding procedure analogous to the approximation algorithms of the popular MAX-CUT problem. As an example, one of the noise models we considered in Cucuringu (2016) is an Erdős-Rényi Outliers model, abbreviated by ERO$(n, p, \eta)$, where the available measurements are given by the following mixture

$$C_{ij} = \begin{cases} r_i - r_j, & \text{with probability } (1-\eta)p \\ \sim \mathcal{U}\;[-(n-1), n-1], & \text{with probability } \eta p \\ 0, & \text{with probability } 1-p, \end{cases} \quad (1)$$

where $r_i$ denotes the unknown ground truth rank of team $i$, $n$ the number of teams, $p$ the probability that a pair of teams play a match against each other, and $\eta$ is the noise level.

The remainder of this section details our synchronization-based ranking algorithm, as well as a number of other state-of-the art methods from the literature. We briefly summarize the Serial-Rank algorithm recently introduced in Fogel et al. (2014), which performs spectral ranking via seriation, and was shown to compare favorably to other classical ranking methods. We also discuss the Rank-Centrality algorithm proposed by Negahban et al. (2012), in the context of rank aggregation. Finally, we consider two other approaches for obtaining a global ranking based on Singular Value Decomposition (SVD) and the popular method of Least Squares (LS). For ease of reference, we summarize in Table 1 the various approaches detailed in the rest of this section.

| Acronym | Name | Section |
|---|---|---|
| SYNC-EIG | Synchronization-Ranking via the spectral relaxation | Sec. 2.1 |
| SER | Serial-Ranking | Sec. 2.2 |
| RC | Rank-Centrality | Sec. 2.3 |
| SVD | SVD Ranking | Sec. 2.4 |
| LS | Least Squares Ranking | Sec. 2.5 |

Table 1: Names of the algorithms we compare, their acronyms, and respective Sections.

|                    | RPS estimate | RPS standard error | ACC estimate | ACC standard error |
|--------------------|--------------|--------------------|--------------|--------------------|
| LS-ord             | 0.2137       | 0.0025             | 0.4676       | 0.0068             |
| RC-ord             | 0.2138       | 0.0025             | 0.4673       | 0.0068             |
| SYNC_EIG-ord       | 0.2144       | 0.0025             | 0.4676       | 0.0068             |
| SYNC_EIG-card      | 0.2147       | 0.0024             | 0.4607       | 0.0068             |
| LS-card            | 0.2148       | 0.0024             | 0.4604       | 0.0068             |
| RC-card            | 0.2150       | 0.0024             | 0.4628       | 0.0069             |
| SYNC_EIG_Sup-card  | 0.2151       | 0.0024             | 0.4629       | 0.0068             |
| SYNC_EIG_Sup-ord   | 0.2152       | 0.0024             | 0.4632       | 0.0068             |
| SVD-ord            | 0.2167       | 0.0024             | 0.4627       | 0.0068             |
| SVD-card           | 0.2168       | 0.0024             | 0.4585       | 0.0068             |
| SER-card           | 0.2173       | 0.0024             | 0.4570       | 0.0067             |
| SER_GLM-card       | 0.2173       | 0.0024             | 0.4570       | 0.0067             |
| SER-ord            | 0.2176       | 0.0024             | 0.4569       | 0.0067             |
| SER_GLM-ord        | 0.2181       | 0.0024             | 0.4576       | 0.0068             |

Table 2: Numerical results obtained by each ranking algorithm, when the resulting features were used on their own in the LF Bradley-Terry formulation.

## 2.1 Sync-Rank: Robust ranking via eigenvector and SDP synchronization

Cucuringu (2016) considered the problem of ranking with noisy incomplete information and made an explicit connection with the angular synchronization problem, for which spectral and SDP relaxations already exist in the literature with provable guarantees. This approach leads to a computationally efficient (as is the case for the spectral relaxation), non-iterative algorithm that is model independent and relies exclusively on the available data.

**The group synchronization problem**

Finding group elements from noisy measurements of their ratios is known as the *group synchronization* problem. It can be applied in settings where the underlying problem exhibits a group structure and one has observed noisy measurements of ratios of group elements. For example, the synchronization problem over the special orthogonal group $SO(d)$ consists of estimating a set of $n$ unknown $d \times d$ matrices $R_1, \ldots, R_n \in SO(d)$ from noisy measurements $R_{ij}$ of a subset of their pairwise ratios $R_i^{-1} R_j$. The least squares solution to synchronization aims to minimize the sum of squared deviations

$$\underset{R_1,\ldots,R_n \in SO(d)}{\text{minimize}} \sum_{(i,j) \in E} w_{ij} \| R_i^{-1} R_j - R_{ij} \|_F^2,$$

where $\|\cdot\|_F$ denotes the Frobenius norm, and $w_{ij}$ are non-negative weights representing the confidence in the noisy pairwise measurements $R_{ij}$. Singer (2011) proposed spectral and semidefinite programming (SDP) relaxations for solving an instance of the above synchronization problem in the context of angular synchronization, over the group SO(2) of planar rotations, where the goal is to estimate $n$ unknown angles

$$\theta_1, \ldots, \theta_n \in [0, 2\pi),$$

given $m$ noisy measurements $\Theta_{ij}$ of their offsets

$$\Theta_{ij} = \theta_i - \theta_j \mod 2\pi.$$

The challenges stem from the amount of noise in the offset measurements, and from the fact that $m \ll \binom{n}{2}$, i.e. only a very small subset of all possible pairwise offsets are measured. In general, one may consider other

4

groups $\mathcal{G}$ (such as SO($d$), O($d$)) for which there are available noisy measurements $g_{ij}$ of ratios between the group elements

$$g_{ij} = g_i g_j^{-1}, \ g_i, g_j \in \mathcal{G}.$$

The set $E$ of pairs $(i,j)$ for which a ratio of group elements is available can be realized as the edge set of a graph $G = (V, E)$, $|V| = n, |E| = m$, with vertices corresponding to the group elements $g_1, \ldots, g_n$, and edges to the available pairwise measurements $g_{ij} = g_i g_j^{-1}$. For the case of angular synchronization (when $\mathcal{G} = \text{SO}(2)$), we start by building the $n \times n$ sparse Hermitian matrix $H = (H_{ij})$ whose elements are either zero or points that lie on the unit circle in the complex plane

$$H_{ij} = \begin{cases} e^{\iota \theta_{ij}} & \text{if } (i,j) \in E \\ 0 & \text{if } (i,j) \notin E. \end{cases} \tag{2}$$

In order to preserve the angle offsets as best as possible, one aims to solve the optimization problem

$$\underset{\theta_1, \ldots, \theta_n \in [0, 2\pi)}{\text{maximize}} \sum_{i,j=1}^{n} e^{-\iota \theta_i} H_{ij} e^{\iota \theta_j} \ , \tag{3}$$

which is incremented by $+1$ whenever an assignment of angles $\theta_i$ and $\theta_j$ perfectly satisfies the given edge constraint $\Theta_{ij} = \theta_i - \theta_j \mod 2\pi$ (i.e. for a *good* edge). The contribution of an incorrect assignment (i.e. of a *bad* edge) is uniformly distributed on the unit circle in the complex plane.

Since (3) is a non-convex and computationally difficult optimization problem (Zhang and Huang, 2006), an alternative is to consider the spectral relaxation

$$\underset{z_1, \ldots, z_n \in \mathbb{C}; \ \sum_{i=1}^{n} |z_i|^2 = n}{\text{maximize}} \sum_{i,j=1}^{n} \bar{z}_i H_{ij} z_j \tag{4}$$

by replacing the individual constraints $z_i = e^{\iota \theta_i}$ having unit magnitude by the much weaker single constraint $\sum_{i=1}^{n} |z_i|^2 = n$. The resulting maximization problem in (4) amounts to maximizing a quadratic form whose solution is known to be given by the top eigenvector of the Hermitian matrix $H$, which has an orthonormal basis over $\mathbb{C}^n$, with real eigenvalues $\lambda_1 \geq \lambda_2 \geq \ldots \geq \lambda_n$ and corresponding eigenvectors $v_1, v_2, \ldots, v_n$. In other words, the spectral relaxation of the non-convex optimization problem in (3) is given by

$$\underset{||z||^2 = n}{\text{maximize}} \ z^* H z, \tag{5}$$

which can be solved via a simple eigenvector computation, by setting $z = v_1$, where $v_1$ is the top eigenvector of $H$, satisfying $Hv_1 = \lambda_1 v_1$, with $||v_1||^2 = n$, corresponding to the largest eigenvalue $\lambda_1$.

As a final step, prior to extracting the final estimated angles, we normalize $H$ by using the diagonal matrix $D$, whose diagonal elements are given by $D_{ii} = \sum_{j=1}^{N} |H_{ij}|$, and define

$$\mathcal{H} = D^{-1} H \ ,$$

which is similar to the Hermitian matrix $D^{-1/2} H D^{-1/2}$, since

$$\mathcal{H} = D^{-1/2}(D^{-1/2} H D^{-1/2}) D^{1/2} \ .$$

Thus, $\mathcal{H}$ has $n$ real eigenvalues $\lambda_1^{\mathcal{H}} > \lambda_2^{\mathcal{H}} \geq \cdots \geq \lambda_n^{\mathcal{H}}$ with corresponding $n$ orthogonal (complex valued) eigenvectors $v_1^{\mathcal{H}}, \ldots, v_n^{\mathcal{H}}$, with $\mathcal{H} v_i^{\mathcal{H}} = \lambda_i^{\mathcal{H}} v_i^{\mathcal{H}}$. We define the estimated rotation angles $\hat{\theta}_1, \ldots, \hat{\theta}_n$ using the top eigenvector $v_1^{\mathcal{H}}$ via

$$e^{\iota \hat{\theta}_i} = \frac{v_1^{\mathcal{H}}(i)}{|v_1^{\mathcal{H}}(i)|}, \quad i = 1, 2, \ldots, n \ .$$

We remark that the estimation of the rotation angles $\theta_1, \ldots, \theta_n$ is unique up to an additive phase since $e^{i\alpha} v_1^{\mathcal{H}}$ is also an eigenvector of $\mathcal{H}$ for any $\alpha \in \mathbb{R}$. This motivates the final post-processing step in our proposed algorithm (Algorithm 1), in which we remove the best circular permutation, as depicted in Figure 2.

**Ranking via angular synchronization**

Let us denote the true ranking of the $n$ teams by $r_1 < r_2 < \ldots < r_n$, and assume without loss of generality that $r_i = i$, i.e. the rank of the $i^{th}$ team is $i$. In the absence of noise, the ranks can be imagined to lie on a one-dimensional line, sorted from 1 to $n$, with the pairwise rank comparisons given, in the noiseless case, by $C_{ij} = r_i - r_j$ (for cardinal measurements) or $C_{ij} = \text{sign}(r_i - r_j)$ (for ordinal measurements). In the angular embedding space, we consider the ranks of the teams mapped to the unit circle, say fixing $r_1$ to have a zero angle with the $x$-axis, and the last team $r_n$ corresponding to an angle equal to $\pi$. In other words, we imagine the $n$ team wrapped around a fraction of the circle, interpret the available rank-offset measurements as angle-offsets in the angular space, and thus arrive at the setup of the angular synchronization problem previously described.

The modulus used to wrap the teams around the circle plays an important role in the recovery process. Choosing to map the teams across the entire circle would cause ambiguity at the end points, since the very highly ranked teams would be positioned very close to (or perhaps even mixed with) the very poorly ranked teams. To avoid this issue, we simply choose to map the $n$ teams to the upper half of the unit circle $[0, \pi]$. This mapping is, in essence, our approach to making the problem of dealing with a non-compact group, such as the real line, amenable to a group synchronization approach. This way, the line is *"compactified"* by simply mapping it to the unit circle (or part of it), making the approach amenable to synchronization methods and less sensitive to outliers.

As previously discussed, since the solution to the angular synchronization problem is computed up to a global shift, one needs to perform an additional post-processing step to accurately extract the ordering of the teams that best matches the given data. For example, as shown in the right plot of Figure 2, Chelsea, Tottenham and Manchester City would lie at the bottom of the ranking (right after Sunderland), while in fact they were the highest ranked teams. To this end, we mod out the best circular permutation of the initial rankings obtained from synchronization, that minimizes the number of upsets in the given data. To measure the accuracy of each candidate circular permutation $\sigma$, we first compute the pairwise rank offsets associated with the induced ranking via

$$P_\sigma(\boldsymbol{s}) = (\sigma(\boldsymbol{s}) \otimes \boldsymbol{1} - \boldsymbol{1} \otimes \sigma(\boldsymbol{s})) \circ A \;, \tag{6}$$

where $\otimes$ denotes the outer product of two vectors $x \otimes y = xy^T$, $\circ$ denotes the Hadamard product of two matrices (entrywise product), and $A$ is the adjacency matrix of the measurement graph $G$. We summarize in Algorithm 1 the main steps of the Sync-Rank algorithm.

Finally, we remark that, throughout the numerical experiments, we only relied on the spectral relaxation, and did not experiment with the semidefinite-programming relaxation. Our current implementation in CVX is computationally costly for large data sets, however the SDP program could be solved efficiently via a Burer and Monteiro (2003) approach, whose surprisingly good empirical performance has only recently been understood theoretically (Boumal et al., 2016).

## 2.2 Serial rank and generalized linear models

Fogel et al. (2014) proposed a seriation algorithm for ranking a set of teams given noisy incomplete pairwise comparisons. Their approach starts by assigning similar rankings to teams that compare similarly with all other teams. The intuition is that teams that beat the same teams and are beaten by the same teams should have a similar ranking in the final solution. They do so by constructing a similarity matrix from the available pairwise comparisons, relying on existing seriation methods to reorder the similarity matrix and thus recover the final rankings.

They make an explicit connection with another related classical ordering problem, namely *seriation*, where one is given a similarity matrix between a set of $n$ items under the assumption that the items have an underlying ordering on the line, such that the similarity between items decreases with their distance. A spectral algorithm that exactly solves the noiseless seriation problem was proposed by Atkins et al. (1998), derived from the observation that, for a given similarity matrix computed from such serial variables, the ordering induced by the second eigenvector of the associated Laplacian matrix (i.e. the *Fiedler* vector) matches that of the variables. Fogel et al. (2014) adapted the above seriation procedure to the ranking problem, and

**Algorithm 1** Summary of the Synchronization-Ranking (Sync-Rank) Algorithm. The gist of the approach can be described as: (1) make the ansatz that the teams are embedded on the upper half of the unit circle, (2) map the resulting goal differentials between pairs of teams to an angle offset in $[0, \pi)$, (3) solve the resulting angular synchronization problem (via the spectral relaxation) that amounts to finding an assignment of angles that best match the given angle offsets (4) mod out the best circular permutation by choosing the ordering that minimizes the number of upsets.

**Input:**
    $G = (V, E)$ the graph of pairwise comparisons.
    $C$ the $n \times n$ matrix of pairwise comparisons (rank offsets), such that whenever $(i, j) \in E(G)$ we have available a (perhaps noisy) comparison between players $i$ and $j$, either a cardinal comparison ($C_{ij} \in [-(n-1), (n-1)]$) or an ordinal comparison $C_{ij} = \pm 1$.

**Output:** Final ranking $r$

1: $r \leftarrow$ Sync-Rank()
Map all rank offsets $C_{ij}$ to an angle $\Theta_{ij} \in [0, 2\pi\delta)$ with $\delta \in [0, 1)$, using the transformation

$$C_{ij} \mapsto \Theta_{ij} := 2\pi\delta \frac{C_{ij}}{n-1}.$$

We choose $\delta = \frac{1}{2}$, and hence $\Theta_{ij} := \pi \frac{C_{ij}}{n-1}$.
Build the $n \times n$ Hermitian matrix $H$ with $H_{ij} = e^{i\theta_{ij}}$, if $(i, j) \in E$, and $H_{ij} = 0$ otherwise, as in (2).
Solve the angular synchronization problem via either its spectral (5) relaxation, and denote the recovered solution by $\hat{r}_i = e^{i\hat{\theta}_i} = \frac{v_1^R(i)}{|v_1^R(i)|}$, $i = 1, 2, \ldots, n$, where $v_1$ denotes the recovered eigenvector.
Extract the corresponding set of angles $\hat{\theta}_1, \ldots, \hat{\theta}_n \in [0, 2\pi)$ from $\hat{r}_1, \ldots, \hat{r}_n$.
Order the set of angles $\hat{\theta}_1, \ldots, \hat{\theta}_n$ in increasing order, and denote the induced ordering by $\boldsymbol{s} = s_1, \ldots, s_n$.
Compute the best circular permutation $\sigma$ of the above ordering $\boldsymbol{s}$ that minimizes the resulting number of upsets with respect to the initial rank comparisons given by $C$

$$r = \underset{\sigma_1, \ldots, \sigma_n}{\arg\min} \ ||\text{sign}(P_{\sigma_i}(\boldsymbol{s})) - \text{sign}(C)||_1$$

with $P$ defined as in (6).

2: $r \leftarrow$ Output as a final solution the ranking induced by the circular permutation $\sigma$.

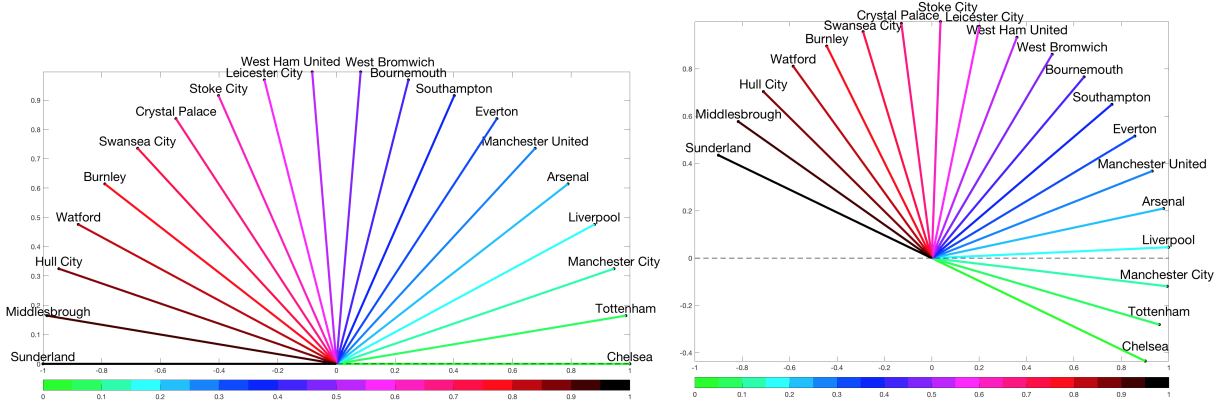

Figure 2: (a) Equidistant mapping of the ranked teams $1, \ldots, n$ around half a circle, for $n = 20$, where the rank of the $i^{th}$ team is $i$ (the ranking used as an example is actually the final standing in the Premier League 2016-2017 season, with Chelsea being ranked first, and Sunderland last). (b) The recovered solution at some random rotation, motivating the step that computes the best circular permutation of the recovered rankings, chosen to minimize the number of upsets with respect to the initially given pairwise measurements.

proposed an efficient polynomial-time algorithm with provable recovery and robustness guarantees, which under certain conditions, is able to perfectly recover the underlying true ranking, even when a fraction of the comparisons are either corrupted by noise or completely missing.

In the case of ordinal measurements (only win-lose information), the proposed similarity measure counts the number of *matching comparisons*. For a given skew symmetric matrix $C$ of size $n \times n$ of pairwise comparisons $C_{ij} = \{-1, 0, 1\}$ (denoting lose, tie or a win), with $C_{ij} = -C_{ji}$, given by the following model

$$C_{ij} = \left\{ \begin{array}{rl} 1 & \text{if team i won over team j,} \\ 0 & \text{if the game ended in a draw,} \\ -1 & \text{if team i lost to team j.} \end{array} \right.$$

Assuming the diagonal of $C$ is set to $C_{ii} = 1, \forall i = 1, 2, \ldots, n$, the similarity matrix takes the form

$$S_{ij}^{match} = \sum_{k=1}^{n} \left( \frac{1 + C_{ik} C_{jk}}{2} \right), \tag{7}$$

where $C_{ik} C_{jk} = 1$ whenever $i$ and $j$ have the same signs, and $C_{ik} C_{jk} = -1$ whenever they have opposite signs. In other words, the similarity $S_{ij}^{match}$ counts the number of matching comparisons between $i$ and $j$ with a third reference team $k$. Written in a compact form, the final similarity matrix is given by

$$S^{match} = \frac{1}{2} \left( n \mathbf{1}\mathbf{1}^T + CC^T \right).$$

The final ranking is the one induced by the Fiedler vector of $S$. The main steps of Serial-Rank algorithm are summarized in Algorithm 2.

Fogel et al. (2014) also consider a generalized linear model setting, where one assumes that the paired comparisons are generated according to a generalized linear model, are independent, and team $i$ defeated team $j$ with probability

$$P_{ij} = H(\nu_i - \nu_j),$$

where $\nu \in \mathbb{R}^n$ is a vector denoting the strength, rank, or skill level of the $n$ teams. They propose the following similarity matrix

$$S_{i,j}^{glm} = \sum_{k=1}^{n} \mathbf{1}_{\{m_{i,k} m_{j,k} > 0\}} \left( 1 - \frac{|C_{i,k} - C_{j,k}|}{2} \right) + \frac{\mathbf{1}_{\{m_{i,k} m_{j,k} = 0\}}}{2}, \tag{8}$$

8

where $m_{i,k} = 1$ if $i$ and $j$ played in a match, and 0 otherwise. The matrix $Q$ of corresponding empirical probabilities is given by the following mixture

$$Q_{i,j} = \begin{cases} \frac{1}{m_{i,j}} \sum_{s=1}^{m_{i,j}} \frac{C_{i,j}^s + 1}{2} & \text{if } m_{i,j} > 0, \\ \frac{1}{2} & \text{if } m_{i,j} = 0. \end{cases}$$

Here, $m_{ij}$ denotes the number of times teams $i$ and $j$ played against each others, and $C_{i,j}^s \in \{-1, 1\}$ is the result of match $s$. We denote by SER-GLM the Serial-Rank algorithm based on the above GLM model given by (8).

---

**Algorithm 2** Serial-Rank: spectral ranking via seriation (Fogel et al., 2014)

**Input:**
  A set of pairwise comparisons $C_{ij} \in \{-1, 0, 1\}$ or [-1,1]
**Output:** Final ranking $r$
  Compute a similarity matrix as shown in (7).
  Compute the associated graph Laplacian matrix

$$L_S = D - S,$$

  where D is a diagonal matrix $D = \mathbf{diag}\,(S\mathbf{1})$, i.e. $D_{ii} = \sum_{j=1}^n G_{i,j}$ is the degree of node $i$ in the measurement graph $G$.
  Compute the Fiedler vector of S (eigenvector corresponding to the smallest nonzero eigenvalue of $L_S$).
  Output the ranking induced by sorting the Fiedler vector of $S$, with the global ordering (increasing or decreasing order) chosen such that the number of upsets is minimized.

---

## 2.3 The rank-centrality algorithm

The third ranking algorithm we consider is Rank-Centrality, introduced by Negahban et al. (2012), and is an iterative algorithm proposed for the rank aggregation problem of integrating ranking information from multiple ranking systems, by estimating scores for the items from the stationary distribution of a certain random walk on the graph of items, where each edge encodes the outcome of pairwise comparisons.

For a pair of teams $i$ and $j$, we let $Y_{ij}^{(l)}$ be equal to 1 if team $j$ beats team $i$, and 0 otherwise, during the $l^{th}$ match between the two teams, for $l = 1, \ldots, k$.

assumes that $\mathbb{P}(Y_{ij}^{(l)}) = \frac{w_j}{w_i + w_j}$, where $w$ represent the underlying vector of positive real weights associated to each player.

Motivated by the Bradley-Terry model (see Section 3 in the main text), Negahban et al. (2012) start by estimating the fraction of times team $i$ has defeated team $j$, and denote this by

$$a_{ij} = \frac{1}{k} \sum_{l=1}^{k} y_{ijl},$$

as long as teams $i$ and $j$ competed in at least one match, and 0 otherwise, where $y_{ijl} = 1$ if team $i$ beat team $j$ in the $l^{\text{th}}$ encounter between the two teams, and 0 otherwise. As a next step, they build the symmetric matrix

$$A_{ij} = \frac{a_{ij}}{a_{ij} + a_{ji}},$$

which converges to $\frac{\pi_i}{\pi_j + \pi_i}$, as $k \to \infty$, where $\pi_i$ represents the 'strength' of team $i$. To define a valid transition probability matrix, one scales all the edge weights by $1/d_{max}$ and considers the resulting random walk

$$P_{ij} = \begin{cases} \frac{1}{d_{max}} A_{ij} & \text{if } i \neq j \\ 1 - \frac{1}{d_{max}} \sum_{k \neq i} A_{ik} & \text{if } i = j, \end{cases}$$

where $d_{max}$ denotes the maximum out-degree of a node, such that each row of $P$ sums to 1. We recover the final rankings by sorting the entries in the corresponding stationary distribution, given by the top left eigenvector of $P$.

## 2.4 Ranking via singular value decomposition

The fourth ranking algorithm we rely on is based on the traditional Singular Value Decomposition (SVD), and was considered in Cucuringu (2016). What makes SVD applicable in this setting is the observation that, in the case of cardinal measurements $C_{ij} = r_i - r_j$, the noiseless matrix of rank offsets $C$ is a skew-symmetric matrix of even rank 2 since

$$R = r\boldsymbol{e}^T - \boldsymbol{e}r^T,$$

where $\boldsymbol{e}$ denotes the all-ones column vector. For a noisy problem, $C$ is a random perturbation of a rank-2 matrix, which motivates us to consider its top two singular vectors, order their entries by their size, extract the resulting rankings, and choose between the first and second singular vector based on whichever one minimizes the number of upsets. Note that since the singular vectors are obtained up to a global sign, we choose the ordering which minimizes the number of upsets.

For the Erdős-Rényi Outliers ERO$(n, p, \eta)$ model (1), the following decomposition could render the SVD-Rank method amenable to a theoretical analysis using tools from the matrix perturbation and random matrix theory literature on rank-2 deformations of random matrices. The expected value of the entries of $C$ is given by

$$\mathbb{E}C_{ij} = (r_i - r_j)(1 - \eta)p,$$

thus $\mathbb{E}C$ is a rank-2 skew-symmetric matrix

$$\mathbb{E}C = (1 - \eta)p(r\boldsymbol{e}^T - \boldsymbol{e}r^T).$$

The decomposition of the given data matrix $C$ into

$$C = \mathbb{E}C + R,$$

where $R = C - \mathbb{E}C$ is a random skew-symmetric matrix whose elements have zero mean makes this approach amenable to a robustness analysis using tools from random matrix theory, in particular low-rank perturbations of large random matrices.

## 2.5 Ranking via least squares

Finally, we also recover rankings via a more traditional least-squares approach. Assuming the number of edges in $G$ is given by $m = |E(G)|$, we denote by $B$ the edge-vertex incidence matrix of size $m \times n$ whose entries are given by

$$B_{ij} = \begin{cases} 1 & \text{if } (i, j) \in E(G), \quad \text{and } i > j \\ -1 & \text{if } (i, j) \in E(G), \quad \text{and } i < j \\ 0 & \text{if } (i, j) \notin E(G), \end{cases}$$

and by $y$ the vector of length $m \times 1$ which contains the pairwise rank measurements $y(e) = C_{ij}$, for all edges $e = (i, j) \in E(G)$. The least-squares solution to the ranking problem is obtained by solving

$$\underset{x \in \mathbb{R}^n}{\text{minimize}} \quad ||Bx - y||_2^2,$$

where $L$ is an array of size $m \times n$, $m$ is the number of edges, $L(e, i) = 1$ and $L(e, j) = -1$ whenever edge number $e$ connects nodes $i$ and $j$, and $b(e) = W(i, j)$ holds the rank offset.



\end{document}